\documentclass[final,5p,times,twocolumn]{elsarticle}

\usepackage{amssymb}
\usepackage{amsmath}
\usepackage{siunitx}
\usepackage{microtype}

\journal{Digital Signal Processing}

\begin{document}
	
	\begin{frontmatter}
		
		
		
		\title{Correct Estimation of Higher-Order Spectra: From Theoretical Challenges to Practical Multi-Channel Implementation in SignalSnap\tnoteref{label1}}
		\author[inst1]{Markus Sifft\corref{cor1}}
\ead{markus.sifft@rub.de}
\author[inst1]{Armin Ghorbanietemad}
\author[inst1]{Fabian Wagner}
\author[inst1]{Daniel Hägele}

\cortext[cor1]{Corresponding author}

\affiliation[inst1]{
	organization={Ruhr University Bochum, Faculty of Physics and Astronomy, Experimental Physics VI (AG)}, 
	addressline={Universitätsstraße 150}, 
	city={Bochum},
	postcode={D-44780}, 
	country={Germany}
}

\title{}

		\begin{abstract}
			Higher-order spectra (Brillinger's polyspectra) offer powerful methods for solving critical problems in signal processing and data analysis. Despite their significant potential, their practical use has remained limited due to unresolved mathematical issues in spectral estimation, including the absence of unbiased and consistent estimators and the high computational cost associated with evaluating multidimensional spectra. Consequently, existing tools frequently produce artifacts—no existing software library correctly implements Brillinger’s cumulant-based trispectrum—or fail to scale effectively to real-world data volumes, leaving crucial applications like multi-detector spectral analysis largely unexplored.
			
			In this paper, we revisit higher-order spectra from a modern perspective, addressing the root causes of their historical underuse. We reformulate higher-order spectral estimation using recently derived multivariate k-statistics, yielding unbiased and consistent estimators that eliminate spurious artifacts and precisely align with Brillinger’s theoretical definitions. Our methodology covers single- and multi-channel spectral analysis up to the bispectrum (third order) and trispectrum (fourth order), enabling robust investigations of inter-frequency coupling, non-Gaussian behavior, and time-reversal symmetry breaking. Additionally, we introduce quasi-polyspectra to uncover non-stationary, time-dependent higher-order features. We implement these new estimators in SignalSnap, an open-source GPU-accelerated library capable of efficiently analyzing datasets exceeding hundreds of gigabytes within minutes.
	 
In applications such as continuous quantum measurements, SignalSnap's rigorous estimators enable precise quantitative matching between experimental data and theoretical models. With detailed derivations and illustrative examples, this work provides the theoretical and computational foundation necessary for establishing higher-order spectra as a reliable, standard tool in modern signal analysis.



		\end{abstract}
		
		
		\begin{highlights}
			\item SignalSnap is the first library for fast, unbiased estimation of polyspectra.
			\item We relate finite-resolution polyspectra to Brillinger’s ideal spectra.
			\item SignalSnap provides for cross-correlation spectra of up to four channels.
			\item We provide single- and multichannel examples for polyspectra.
			\item Symmetries and isomorphisms of multi-channel polyspectra are derived.
			
		\end{highlights}
		
		
	\end{frontmatter}
	
		
		
		
		\section{Introduction}
			\label{sec:introduction}
			
		In 2003, Birkelund and colleagues expressed a frustration that remains relevant today:
		\begin{quote}
			“In theory, polyspectra can be applied to solve many important problems in signal processing and data analysis. In practice, however, one has been discouraged by the poor statistical properties of most polyspectral estimators.”~\cite{birkelundSP2003}
		\end{quote}
		
		This observation highlights a long-standing gap between the theoretical promise and practical use of higher-order spectral analysis. The bispectrum, trispectrum, and higher-order generalizations provide access to non-Gaussian behavior, time-reversal symmetry breaking, and complex inter-frequency correlations that cannot be captured by the power spectrum alone. Yet, despite their value, higher-order spectra remain underused across many fields of science and engineering.
		
		In fact, we observe two fundamental problems with the current state of literature and available software implementations for higher-order spectral analysis. The first concerns the definition of higher-order spectra itself: Brillinger's definitions of polyspectra 		$S_z^{(n)}(\omega_1,...,\omega_{n-1})$
		\begin{align}
			C_n(z(\omega_1),..., z(\omega_n))   =   2\pi \delta(\omega_1+...+\omega_n)S_z^{(n)}(\omega_1,...,\omega_{n-1}). 	  \label{eq:defPolyspectra}
		\end{align}
		are based on higher-order cumulants, which are crucial for distinguishing true non-Gaussian from Gaussian contributions and allow for the simple subtraction of background noise spectra \cite{Brillinger1965}. While for second- and third-order spectra the moment-based and cumulant-based formulations are equivalent for average-free signals, this equivalence breaks down at fourth order. None of the existing libraries we reviewed correctly implement the trispectrum using this cumulant-based definition. As a result, moment-based trispectra may include additional false structures such a an offset or other artifacts. We examine this issue in detail in Section~\ref{sec:HigherOrderSpectra}, where we show that these methods yield significant non-zero trispectra even for white Gaussian noise (see Figure~\ref{fig:hosa_comparison}) contradicting the expected theoretical outcome.
		
		The second, closely related issue is the \textit{estimation} of polyspectra from finite datasets. Estimating the power spectrum involves the variance, which can be handled using well-understood, unbiased estimators. In contrast, Brillinger's higher-order spectra depend on third- and fourth-order cumulants whose unbiased estimation from finite data is significantly more subtle and often mishandled in practice \cite{Brillinger1965}. All implementations known to us rely on \textit{biased} or only \textit{asymptotically unbiased} estimators \cite{trappMSSP2021}. This distinction is crucial: a biased estimator does not converge to the correct value for any finite $m$, which can lead to systematic artifacts in the resulting spectrum.
		
		For example, consider the second-order cumulant $C_2(x, x) = \langle x^2 \rangle - \langle x \rangle^2$, where $\langle \dots \rangle$ represents an average over infinitely many samples. A well-known unbiased estimator is
		\begin{equation}
			\label{eq:c2_estimator_intro}
			c_2(x, x) = \frac{m}{m - 1} \left( \overline{x^2} - \overline{x}^2 \right),
		\end{equation}
		where $\overline{(\dots)}$ denotes the sample mean for $m$ samples. The estimator includes the famous prefactor $(m-1)^{-1}$ known as the Bessel correction \cite{kenneyBOOK1951} which ensures $\langle c_2 \rangle = C_2$ \cite{kenneyBOOK1951}. In contrast, the so-called natural estimator
		\begin{equation}
			c_2'(x, x) = \overline{x^2} - \overline{x}^2
		\end{equation}
		is biased, with $\langle c_2' \rangle = C_2 + \mathcal{O}(1/m)$, where $ \mathcal{O}(1/m)$ indicates the order of the error. As $m$ increases, the bias decreases. Nevertheless, for many practical applications where a low $m$ is required, the error remains significant. An extreme case is highlighted in \cite{starosielecRSI2010}, where spectra are computed from averages over just $m=2$ samples, making the use of unbiased estimators indispensable.

		To address the problem of unbiased estimation, we employ the multivariate generalization of Fisher’s $k$-statistics, which provides estimators that are both unbiased and consistent. These estimators correct for finite-sample effects at all relevant orders, and to our knowledge, have not been systematically used in polyspectral analysis before. Their structure parallels that of the cumulants but includes $m$-dependent prefactors, which we previously derived up to fourth order \cite{schefczikARXIV2019}.
		
		Another frequently overlooked issue is the correct \textit{normalization} of the spectral estimates when using windowed data. In practical applications, signal segments are multiplied by window functions to reduce spectral leakage \cite{harrisProc1978}, and it is essential that the spectral estimator accounts for the window length $T$, the number of data points $N$, and the window coefficients $g_i$. Incorrect or missing normalization prevents meaningful comparisons across different datasets or window configurations. In this work, we derive the exact normalization factors for each spectral order and relate them to the Fourier transform of the window function (Section \ref{sec:Estimation2ndOrder} and \ref{sec:HigherOrderSpectra}). This ensures that spectra obtained for different values of $T$, $N$, and $g_i$ remain quantitatively consistent and comparable.
		
		Beyond the estimation problem, computational cost has long discouraged the use of higher-order spectra. While the power spectrum is a one-dimensional quantity that can be efficiently computed, the bispectrum and trispectrum are inherently multidimensional, involving combinations of Fourier coefficients across multiple frequencies and potentially multiple signals. As a result, memory requirements and computation time increase rapidly. For large datasets—as are common in quantum experiments, neurophysiology, and other data-intensive domains—existing tools become impractical. To overcome this, we introduce \texttt{SignalSnap}, an open-source, GPU-accelerated library capable of evaluating multidimensional spectra from hundreds of gigabytes of data in a matter of minutes.
		
		A further limitation in the current literature is the lack of generalization to \textit{multichannel} signals up to fourth order. While experimental setups routinely acquire multichannel data, the formalism for cross-bispectra and cross-trispectra has remained incomplete. We present for the first time a systematic derivation of multichannel spectrum estimators based on $k$-statistics and analyze the resulting symmetries, which are essential for interpreting spectral correlations between subsystems or sensor channels.
		
		In the sections that follow, we present the theoretical foundations, algorithmic implementation, and example applications of our approach. Readers already familiar with the structure and motivation of higher-order spectra may wish to skip directly to Section~\ref{sec:Estimation2ndOrder}, where the derivations begin. For others, the next section provides a compact introduction to higher-order spectra and their practical relevance.

		\section{Motivation and Background on Higher-Order Spectra}

		To fully appreciate the practical and theoretical relevance of higher-order spectra, it is helpful to revisit their foundations and understand where conventional techniques fall short.
		
		The analysis of stochastic signals is a cornerstone across nearly all fields of science and engineering \cite{Brillinger2001, Papoulis2002}. Applications range from audio processing \cite{Oppenheim1999}, financial time series analysis \cite{BlackJPE1973}, and biomolecular dynamics via photon statistics \cite{JaegerCPC2009} to precision measurements in quantum technologies \cite{ClerkRMP2010, hagelePRB2018, sifftPRR2021, SifftPRA2023, SifftPRA2024}. Classical electrical signals, such as voltage fluctuations across resistors, often exhibit characteristic $1/f$ noise linked to underlying material processes \cite{DuttaRMP1981}. In each case, statistical descriptors of the signal encode essential insights into the system’s structure or dynamics.
		
			A common starting point for analyzing a continuous real-valued, time-dependent stochastic signal $z(t)$ is to define and calculate statistical quantities. Here we assume that $z(t)$ is a stationary process, i.e. its statistical properties remain invariant under time shifts. Under this assumption, the mean $m_z = \langle z(t) \rangle$ and variance $\sigma_z^2 = \langle z(t)^2 \rangle - \langle z(t) \rangle^2$ can be defined.
		More refined frequency-domain information is obtained through the power spectrum
		\begin{equation}
			S_z^{(2)}(\omega) \propto \langle z(\omega) z^*(\omega) \rangle + \ldots,
		\end{equation}
		where $z(\omega)$ is the Fourier transform of a single realization of $z(t)$ [see \ref{app:Fourier}], $z^*(\omega)$ denotes its complex conjugate, and the average is taken over an ensemble of realizations. Like the variance, this quantity is of second-order in the signal but now reveals the intensity of the signal at different frequencies and, for stationary processes, has infinite spectral resolution in theory.
		
		However, the power spectrum alone is blind to many important signal characteristics. Notably, it is entirely insensitive to non-Gaussian features which is important since distinct processes with different higher-order statistics may share the same power spectrum. This limitation motivates the use of higher-order spectra—generalizations that capture correlations among multiple frequencies and reveal structure beyond second-order statistics.
		
		\begin{figure*}[t]
			\centering
			\includegraphics[width=\linewidth]{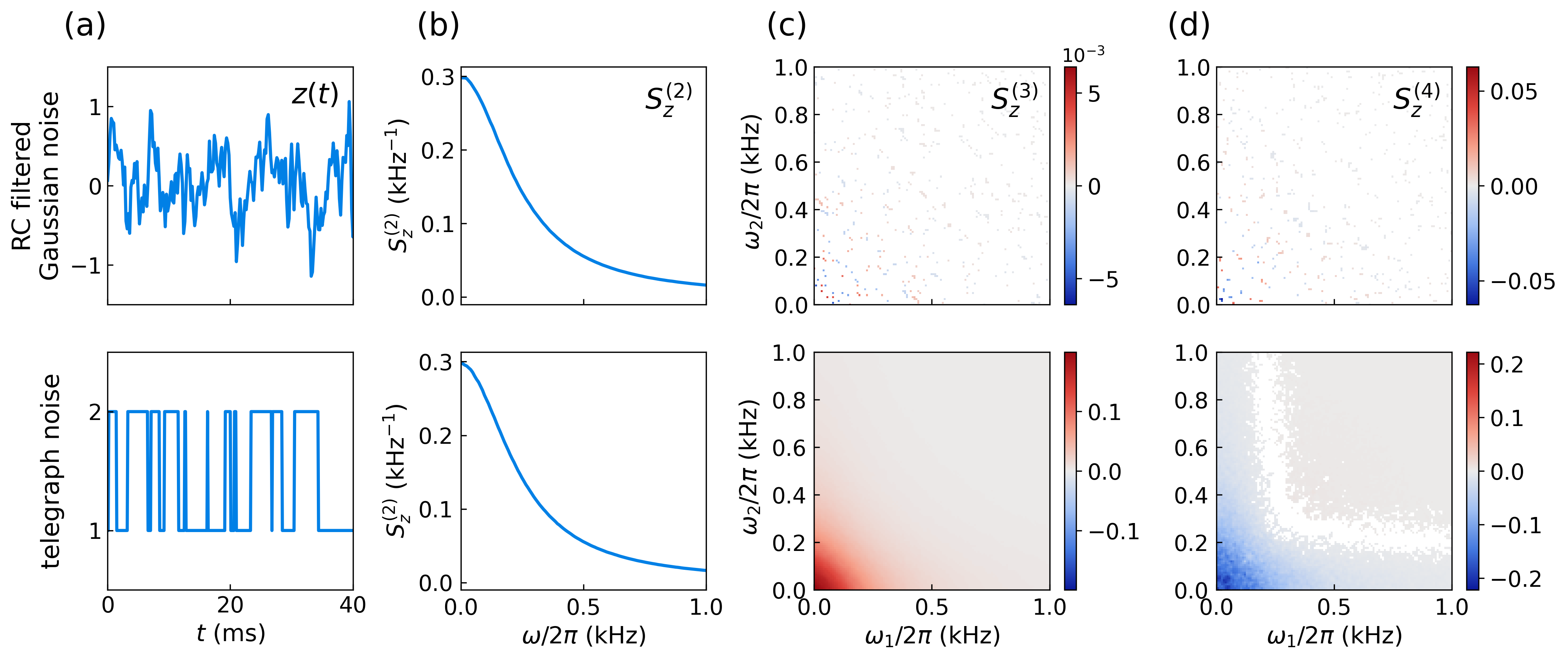}
			\caption{Comparison of RC-filtered white noise (upper row) and random telegraph noise (lower row). (a) Samples of the time-dependent signals are clearly distinct. (b) The power spectra $S_z^{(2)}(\omega)$ are identical and follow a Lorentzian shape. (c) The bispectrum $S_z^{(3)}(\omega_1,\omega_2)$ exhibits no significant structure for RC-filtered white noise but shows a strong positive peak for random telegraph noise. (d) The 
				trispectrum $S_z^{(4)}(\omega_1,\omega_2)$ exhibits a significant structure only for random telegraph noise. 
				Higher-order spectra thus are important for distinguishing signals with identical second-order statistics. Spectral values below the $3\sigma$ noise level are displayed as white points. Spectra $S_z^{(3)}$ and $S_z^{(4)}$ are given in units of kHz$^{-2}$ and kHz$^{-3}$, respectively.}
			\label{fig:intro_plot}   
		\end{figure*}
		
		To illustrate this, we compare two signals with visually distinct time-domain behavior: RC-filtered Gaussian white noise and two-state random telegraph noise [see Figure \ref{fig:intro_plot}(a)]. Despite their different origins, both signals exhibit identical power spectra—simple Lorentzian functions centered at $\omega = 0$. The RC-filtered noise is modeled as
		\begin{equation}
			\label{eq:intro_model}
			\frac{d}{dt} z(t) + \gamma z(t) = \gamma S_0^{1/2} \Gamma(t),
		\end{equation}
		where $\Gamma(t)$ is delta-correlated white noise and $\gamma$ sets the filter time constant. The resulting power spectral density is
		\begin{equation}
			S^{(2)}_{\text{RC}}(\omega) = \frac{S_0}{1 + (\omega/\gamma)^2}.
		\end{equation}
		Similarly, random telegraph noise is modeled as a two-state continuous-time Markov process with transition rates $\gamma_1$ and $\gamma_2$, yielding \cite{sifftPRR2021}
		\begin{equation}
			S^{(2)}_{\text{Telegraph}}(\omega) = \frac{2\gamma_1 \gamma_2}{(\gamma_1 + \gamma_2)^3} \cdot \frac{1}{1 + \omega^2/(\gamma_1 + \gamma_2)^2}.
		\end{equation}
		Both power spectra fully coincide for
		$S_0 = \frac{2\,\gamma_{1}\,\gamma_{2}}{(\gamma_{1}+\gamma_{2})^3}$ and $\gamma = \gamma_{1} + \gamma_{2}$.

		The distinction between these signals emerges only when examining their higher-order spectra. 	Historically, higher-order spectral analysis dates back at least to 1953 \cite{blancBOOK1953}. Brillinger, Mendel, Nikias, and others made major contributions through the 1980s and 1990s \cite{Brillinger1965, mendelIEEE1991, NikiasIEEE1993}. Brillinger introduced so-called polyspectra which generalize the concept of the power spectrum to higher orders. The third-order bispectrum 
		\begin{equation}
			S_z^{(3)}(\omega_1, \omega_2) \propto \left\langle z\left(\omega_1\right) z\left(\omega_2\right) z^*\left(\omega_1+\omega_2\right)\right\rangle+\ldots
		\end{equation}
		and the fourth-order trispectrum
		\begin{equation}
			S_z^{(4)}(\omega_1, \omega_2) \propto 
			\left\langle z^*\left(\omega_1\right) z\left(\omega_1\right) z^*\left(\omega_2\right) z\left(\omega_2\right) \right\rangle + ...
		\end{equation}
		are known to
		reveal non-Gaussian behavior (see e.g. \cite{birkelundSP09,hasselmann1963bispectra}) such as time-reversal symmetry breaking in $S_z^{(3)}$ (see \ref{app:timeinversion}) and correlations between intensity contributions to the signal at different frequencies in $S_z^{(4)}$. The exact definition of polyspectra in terms of cumulants is given in Sec. \ref{sec:HigherOrderSpectra}. 

		Fig.~\ref{fig:intro_plot} shows that the power spectra of RC-filtered white noise and telegraph noise can be identical while their higher-order spectra exhibit distinct differences. The bispectrum and trispectrum of the RC-filtered noise (upper row) exhibit no significant non-zero values. All values that are within their $3\sigma$ error bounds are colored in white. Vanishing higher-order spectra are expected since RC-filtered Gaussian noise remains Gaussian after linear filtering. In contrast, the telegraph noise (lower row) shows significant contributions in both the bispectrum and trispectrum, highlighting its non-Gaussian dynamics. All spectra of Fig.~\ref{fig:intro_plot} were calculated with the SignalSnap library.  We recently exploited polyspectra up to the fourth order of telegraph noise to recover Markov transition rates \cite{sifftPRR2021,SifftPRA2024} even in cases involving hidden Markov dynamics with more than two underlying states \cite{sifft2025PRB}. 
		
Despite their obvious use demonstrated above, the widespread adoption of higher-order spectra has been hindered by biased or inconsistent estimators, high computational demands, and  difficulties in the interpretation of multidimensional spectra. Moreover, early toolboxes - such as HOSA \cite{swami2025hosa} - did not incorporate consistent normalization across different windowing schemes or data lengths. With SignalSnap, we aim to overcome these barriers by providing rigorously derived, unbiased estimators based on multivariate $k$-statistics; fast, GPU-accelerated implementations; correct normalization that accounts for window length, sampling rate, and tapering; and full support for multi-channel generalizations with symmetry analysis. All of these are essential for making higher-order spectra a reliable tool in modern signal analysis.
		
		In the next sections, we begin by formalizing the estimation of second- and higher-order spectra and present the necessary mathematical framework. This will allow us to build up toward general multi-detector expressions and their practical implementation in SignalSnap.

		\section{Estimation of the second-order spectrum}
		\label{sec:Estimation2ndOrder}
		In literature one often finds the power spectrum 
		\begin{equation}
			S_z^{\rm lit}(\omega) = \int_{-\infty}^\infty e^{j \omega \tau} \langle z(t + \tau) z(t) \rangle \, d \tau
		\end{equation}
		defined in terms of the Fourier transform of the autocorrelation of $z(t)$ (Wiener-Khinchin theorem).  
		The autocorrelation \mbox{$\langle z(t + \tau) z(t) \rangle$} is a second-order moment of $z(t)$. If $z(t)$ has a non-zero mean the spectrum $S_z^{\rm lit}(\omega)$ suffers from a delta-function at $\omega = 0$. This can be avoided by replacing the second-order moment with the covariance $C_2(x,y) = \langle x y \rangle -  \langle x  \rangle \langle y \rangle$, which is identical to the second-order cumulant for two variables.
		Hence, we define the second-order spectrum as
		\begin{equation}
			\label{S2_autocorr}
			S_z^{(2)}(\omega) = \int_{-\infty}^\infty e^{j \omega \tau} C_2(z(t + \tau), z(t)) \, d \tau.
		\end{equation}
		Since literature offers definitions of power spectra with varying prefactors or with a dependency 
		on frequency $f$ rather than on $\omega = 2 \pi f$, we 
		quote an important relation for our definition:
		The variance $\sigma^2_z$ of $z(t)$ relates to the spectrum via 
		\begin{align}
			\int_{-\infty}^\infty  S_z^{(2)}(\omega)\, d\omega  &= 2\pi C_2(z(t),z(t)) = 2 \pi \sigma^2_z. 
		\end{align}  
		
		The spectrum $S_z^{(2)}(\omega)$ can be expressed in terms of the Fourier transform
		\begin{equation}
			z(\omega) = \int_{-\infty}^\infty e^{j \omega t} z(t)\, dt, \label{eq:zFourier}
		\end{equation}  
		where we distinguish $z(t)$ and its Fourier transform only by the different arguments $t$ (or $\tau$) and $\omega$ (see \ref{app:Fourier}). 
		One easily finds the relation (see Chapter 1 in Gardiner \cite{gardinerBOOK2009})
		\begin{equation}
			C_2(z(\omega),z(\omega’)) = 2\pi \delta(\omega+\omega’)S_z^{(2)}(\omega) \label{eq:defC2}
		\end{equation}
		or equivalently
		\begin{equation}
			C_2(z(\omega),z^*(\omega’)) = 2\pi \delta(\omega -\omega’)S_z^{(2)}(\omega), \label{eq:C2S2}
		\end{equation}
		since $ z(\omega) = z^*(-\omega)$ for real-valued $z(t)$.
		The delta function is a consequence of the stationarity condition where the autocorrelation depends only on $\tau$ but not on $t$.
		This finding will lead us to a recipe for estimating $S_z^{(2)}(\omega)$ and its higher-order generalizations from data which is based on Fourier coefficients and their cumulants.  This takes advantage of the Fourier coefficients being very efficiently calculated via the fast Fourier transformation algorithm. 
		
		We emphasize that the definition of the power spectrum, Eq. (\ref{eq:defC2}), cannot immediately be applied to its calculation from a data stream $z(t)$. The calculation of the Fourier transformation of $z(t)$ would according to Eq. (\ref{eq:zFourier}) require the knowledge of $z(t)$ in an interval from minus to plus infinity. Moreover, the definition of  $C_2$ assumes that the moments required for finding the cumulant are determined from an infinite number of samples of $z(t)$. In reality, however, $z(t)$ is usually measured once in a finite time interval. In the following, we provide a scheme that is able to find estimates for the power spectrum from a discretely sampled signal of finite length. Such estimates will exhibit stochastic errors and a limited spectral resolution that depends on a temporal window function.
		
\begin{figure}[t]
			\centering
			\includegraphics[width=6.5cm]{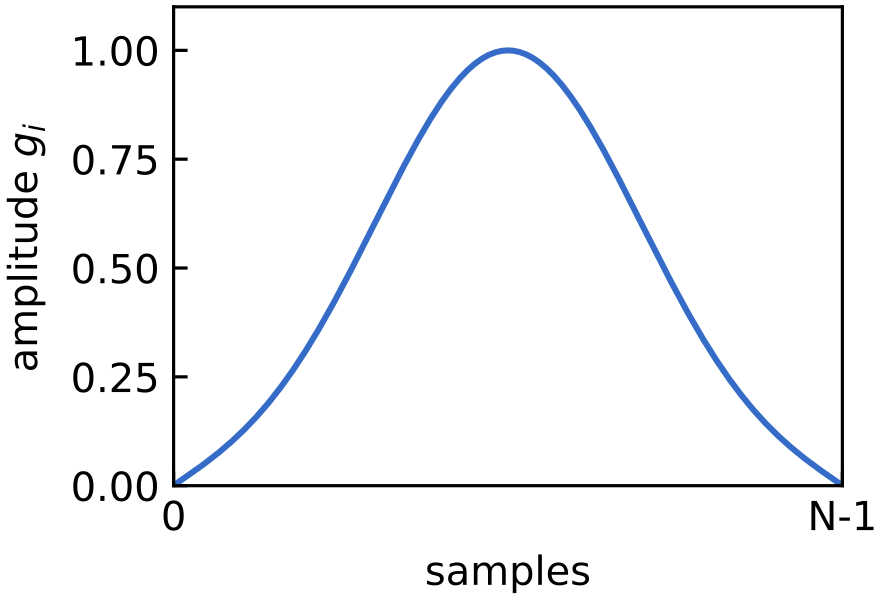}
			\caption{SignalSnap applies the Approximate Confined Gaussian
				Window $g_i$ to the data before calculating the discrete Fourier transform via the fast Fourier transformation. The window parameters are $\sigma_t = 0.14$ for the widths and $T = N$
				for the window length
				(\ref{app:cgw}) \cite{starosielecSP2014}. }
			\label{fig:window_function}   
		\end{figure}

		The Fourier coefficients of $z(t)$ will be calculated via the discrete Fourier transformation in time windows of lengths $T$. We assume that the signal $z(t)$ is known at $N$ equidistant points within the interval. We, therefore, define
		\begin{equation}
			z_i = z(i T/N - t_0),
		\end{equation}  
		where $0 \le i < N-1$ and $t_0$ is the position of the time interval in time.
		We also define a discrete window function 
		\begin{equation}
			g_i = g(i T/N),
		\end{equation}
		which will enter the calculation of Fourier coefficients. Window functions are routinely used in signal processing for improving the spectral resolution \cite{harrisProc1978}. The SignalSnap library uses the approximate confined Gaussian window with window parameter $\sigma_t = 0.14$ (Fig. \ref{fig:window_function}) for its optimal root mean square (RMS) time-bandwidth product \cite{starosielecSP2014} (see \ref{app:cgw}). 
		Parts of the signal that belong to the middle of the window function cause a stronger contribution to the spectra than parts that are outside that region.
		This effect can be reversed by calculating the same spectrum with windows shifted by $T/2$. SignalSnap has the option to calculate both (interlaced) spectra and 
		to display their average. 
		
		The coefficients $a_k$ of the discrete Fourier transformation of $z_i$ are defined as
		\begin{equation}
			\label{eq:a_k}
			a_k = \frac{T}{N} \sum_{i=0}^{N-1} g_i z_i e^{2 \pi i j k / N} e^{-j \frac{2\pi k t_0}{T}}
		\end{equation}
		for $k = 0, 1, \dots, N-1$. The factor $e^{-j \frac{2\pi k t_0}{T}} = e^{-j \omega t_0}$ shifts the signal to account for the window start time $t_0$. However, from  Eq. (\ref{eq:C2S2}), we see that this factor does not affect the power spectrum, because the $\delta$-function ensures that only terms with $\omega - \omega' = 0$ contribute. This conclusion will hold for higher-order spectra as well. Therefore, without loss of generality, we can set $t_0=0$ in Eq. \eqref{eq:a_k} to simplify our expressions. The fast Fourier transformation is applied in SignalSnap for very efficiently calculating all coefficients $a_k$, where $N$ is typically a power of two or at least the product of small primes.
		
		Inspired by Eq. (\ref{eq:C2S2}), we will next derive a relation between cumulants of the coefficients $a_k$ and the power spectrum $S_z^{(2)}(\omega)$.
		The sum on the RHS of Eq. (\ref{eq:a_k}) can be approximated by an integral 
		\begin{equation}
			a'_k = \int_0^T g(t) z(t) \exp(2 \pi j k t / T) \, dt,
		\end{equation} 
		where $a_k \approx a'_k$ for smooth $z(t)$.
		After defining $g(t) = 0$ outside the interval, i.e., for $t<0$ or $t > T$ and introducing $\omega_k = 2 \pi k / T$, we obtain
		\begin{align}
			a'_k & =  \int_{-\infty}^\infty g(t) z(t) e^{j \omega t}  \, dt |_{\omega = \omega_k} \nonumber \\
			& =  [g(\omega) * z(\omega)]_{\omega = \omega_k},
		\end{align}
		where $*$ denotes the convolution (\ref{app:Fourier}).

		We find for the second-order cumulant
		\begin{align}
			C_2(a_k,a^*_k) & \approx    C_2(a'_k,a'^*_k) \nonumber \\
			& =   \frac{1}{(2 \pi)^2} \iint  C_2(z(\omega), z^*(\omega')) \nonumber \\
			&   \times g(\omega_k - \omega) g^*(\omega_k - \omega') \,d\omega\,d\omega'
			\nonumber \\
			& =    \frac{1}{2 \pi}  \int S^{(2)}_z(\omega) g(\omega_k - \omega) g^*(\omega_k - \omega) \,d\omega \nonumber \\
			& =  [ |g(\omega)|^2 * S_z^{(2)}(\omega)]_{\omega = \omega_k}, \label{eq:approx_s2}
		\end{align}
		where we made use of the multilinearity of cumulants in line two  [$C_2(a x, b y) = a b C_2(x,y)$, with constants $a$ and $b$], and used Eq. (\ref{eq:C2S2}) to arrive at the third line.
		
		The RHS of Eq. \eqref{eq:approx_s2} corresponds almost to the ideal spectrum $S^{(2)}_z(\omega)$. The ideal spectral resolution is, however, compromised by the convolution with $|g(\omega)|^2$. Window functions $g$ are in general designed to be spectrally narrow and peaked at $\omega = 0$. We can, therefore, consider the RHS of Eq. (11) to be a good approximation of the spectrum $S^{(2)}_z(\omega)$ apart from a normalization factor that follows from the area $A$ under the peak of $|g(\omega)|^2$. We find
		\begin{align}
			A & =  \int g(\omega) g^*(\omega)\,d\omega \nonumber \\
			& =  \iiint g(t) e^{j \omega t} g^*(t') e^{-j \omega t'} \, dt dt' d\omega \nonumber \\
			& =  2 \pi \int g(t) g^*(t) \, dt \nonumber \\
			& \approx  2 \pi \frac{T}{N} \sum_i g_i g^*_i.
		\end{align}
		Considering that the prefactor $(2 \pi)^{-1}$ of the convolution integral [Eq. \eqref{eq:convolution}] in the frequency domain reduces the effective peak area of $|g(\omega)|^2$,
		we  find the following approximation for the ideal spectrum
		\cite{Schefczik2020}
		\begin{equation}
			S^{(2)}_z(\omega_k) \approx \frac{N C_2(a_k, a_k^*)}{T \sum_{i = 0}^{N-1} g_i g^*_i}
		\end{equation}
		in terms of the Fourier coefficients $a_k$, the temporal window length $T$, the number of support points $N$, and the window function $g_i$.
		The equation above is not yet a complete recipe for calculating an estimate of $S^{(2)}_z(\omega)$ from a data stream $z_i$. In the case of a finite amount of data, the second-order cumulant   $C_2(x,y)$ needs to be estimated with special care to avoid systematic errors. 
		We will discuss properties of cumulant estimators and their generalization to higher orders in paragraph  \ref{para:unbiasedestimators}.

		\section{Higher-order spectra of a single channel}
		\label{sec:HigherOrderSpectra}
		Brillinger’s polyspectra 
		$S_z^{(n)}(\omega_1,...,\omega_{n-1})$  can be defined via higher-order cumulants of $z(\omega)$ \cite{Brillinger1965}
		\begin{align}
			C_n(z(\omega_1),..., z(\omega_n)) &   \nonumber \\
			&  \hspace{-2cm} =   2\pi \delta(\omega_1+...+\omega_n)S_z^{(n)}(\omega_1,...,\omega_{n-1}). 	  \label{eq:defPolyspectra}
		\end{align}
The second order polyspectrum is identical with the power spectrum in Eq. (\ref{eq:defC2}).

		Cumulants can be represented in terms of products of moments as \cite{mendelIEEE1991,gardinerBOOK2009}
		\begin{align}
			C_2(x,y) & =  \langle yx \rangle -\langle y \rangle \langle x \rangle, \label{eq:C2}  \\
			C_3(x,y,z) & =  \langle zyx \rangle - \langle yx \rangle\langle z \rangle \nonumber\\
			&- \langle zx \rangle\langle y \rangle
			- \langle zy \rangle\langle x \rangle + 2 \langle z \rangle\langle y \rangle\langle x \rangle, \label{eq:C3} \\
			C_4(x,y,z,w) & =  \langle wzyx \rangle - \langle wzy \rangle\langle x \rangle - \langle wyx \rangle\langle z \rangle\nonumber \\
			&-\langle wzx \rangle\langle y \rangle-\langle zyx \rangle\langle w \rangle -\langle wz \rangle\langle yx \rangle \nonumber \\ \nonumber
			&-  \langle wy \rangle\langle zx \rangle-\langle wx \rangle\langle zy \rangle +2\langle yx \rangle\langle w \rangle \langle z \rangle \\ \nonumber
			&+2\langle zx \rangle\langle w \rangle \langle y \rangle+2\langle wx \rangle\langle y \rangle \langle z
			\rangle \\ \nonumber
			&+ 2\langle wy \rangle\langle z \rangle \langle x \rangle+2\langle zy \rangle\langle w \rangle \langle x \rangle \\ 
			&+ 2\langle wz \rangle\langle y \rangle \langle x \rangle - 
			6\langle x \rangle\langle y \rangle\langle z \rangle\langle w \rangle. \label{eq:C4}
		\end{align}
		The cumulants above and cumulants of even higher orders are obtained from a cumulant generating function (see e.g. \cite{hagelePRB2018,gardinerBOOK2009}). Wolinsky labeled cumulants as "simply expectations with lower-order dependence removed", which nicely puts the fact, that higher-order cumulants extract additional information about a signal without repeating redundant information contained in lower-order cumulants \cite{wolinsky1988}. 
		We emphasize that the cumulant of the sum of two {\it independent} processes is equal to the sum of their individual cumulants. For two independent stochastic measurement records, $z(t)$ and $w(t)$, it follows that   $S_{z + w}^{(n)}  =  S_z^{(n)} +  S_w^{(n)}$. This property facilitates the subtraction of a background noise spectrum, which can often be measured separately in practice \cite{mendelIEEE1991}.

		Expressions for the third- and fourth-order polyspectra are derived through a 
		calculation similar to that for the power spectrum (see \ref{C3S3})
		\begin{align}
			S^{(3)}_z(\omega_k, \omega_l) &\approx \frac{N C_3(a_k, a_l, a_{k+l}^*)}{T \sum_{i = 0}^{N-1} g_i^2 g^*_i} \\
			S^{(4)}_z(\omega_k, \omega_l,\omega_p) &\approx \frac{N C_4(a_k, a_l, a_p, a_{k+l+p}^*)}{T \sum_{i = 0}^{N-1} g_i^3 g^*_i}.
		\end{align}
		For completeness, we also mention the first-order spectrum, which is a single number (see \ref{C1S1})
		\begin{equation}
			S^{(1)}_z  \approx   \frac{N C_1(a_0)}{T \sum_{i = 0}^{N-1} g_i}, 
		\end{equation}
		where $C_1(x) = \langle x \rangle$. 
		The SignalSnap library implements only a two-dimensional fourth order spectrum by taking a specific plane cut through the full spectrum, defined as
		\begin{equation}
			S^{(4)}_z(\omega_k, \omega_l) \approx \frac{N C_4(a_k, a_k^*, a_l, a_l^*)}{T \sum_{i = 0}^{N-1} g_i^2 (g^*_i)^2},
		\end{equation}
		which can be interpreted as an intensity correlation between two frequencies, $\omega_k$ and $\omega_l$. This limitation is primarily due to computational constraints. A full three-dimensional trispectrum with $10^3$ points per axis would require storing $10^9$ values, which is challenging to manage. Furthermore, we argue that most signals exhibit contributions primarily in that plane. Non-zero contributions outside that plane would require a phase correlation among four frequencies. \ref{app:S4threeD} presents a workaround with which two-dimensional spectra of other parallel planes
		can be calculated.

		The approximate expressions for $S^{(1)}_z$ to $S^{(4)}_z$ include correct normalization factors which depend on the window length $T$, the number of discretization points $N$, and the window coefficients $g_i$. These factors are essential for comparing spectra obtained using varying window functions or different window lengths. Correct prefactors are also vital for comparing experimental spectra with spectra obtained from a theory \cite{sifftPRR2021}. The correct normalization factor for the third-order case was previously addressed by Huber {\it et al.}, but without deriving a connection to the Fourier transformation of the window function \cite{Huber1971}.

		\subsection{Unbiased Cumulant Estimators} 
		\label{para:unbiasedestimators}

		The values of the cumulants that appear in the expressions for the approximate polyspectra $S^{(1)}_z$ to $S^{(4)}_z$ must in the case of limited
		data be estimated with suitable cumulant estimators. 
	 The generalization of $c_2(x,x)$ for one variable and up to sixth order was given by Fisher and is today known as the k-statistics \cite{FisherPLMS1928,kendallBOOK1943}. It is important to note, that the k-statistics requires the samples of $x$ to be independent and identically distributed (i.i.d.) (e.g. the results of throwing dice). Otherwise, averages of an estimator may not converge to the correct cumulant.

		The SignalSnap library uses the multivariate version of the k-statistics to estimate cumulants of the Fourier coefficients [see Eqs. (\ref{eq:C2})-(\ref{eq:C4})].  The k-statistics for two variables has been known before, see \cite{cookBIOMETRIKA1951}, while explicit expressions for three or four variables are hard to find. This may be the reason why we could not find any reference to Fisher's k-statistics in the literature on polyspectra, except for Gardner in \cite{gardnerIEEE1994}, who, however, did not follow up on them. The estimators below were derived and discussed by two of the authors \cite{schefczikARXIV2019}:
		\begin{align}
			c_2(x,y) =& \frac{m}{m-1}\left( \overline{xy} - \overline{x} \,\overline{y} \right), \label{eq:c2_estimator}\\
			c_3(x,y,z) =& \frac{m^2}{(m-1)(m-2)}\nonumber \\
			&\times\left( \overline{xyz}- \overline{xy}\, \overline{z} - \overline{xz}\, \overline{y} \right. \nonumber \\
			&- \left. \overline{yz}\, \overline{x} + 2 \overline{x}\,\overline{y}\, \overline{z} \right), \label{eq:c3_estimator}
		\end{align}
		\begin{align}
			c_4(x,y,z,w) =& \frac{m^2}{(m-1)(m-2)(m-3)} \nonumber \\
			& \nonumber \hspace{-0.4cm} \times \Big[(m+1) \overline{xyzw}\Big.  \\
			& \hspace{-0.4cm}\Big.- (m+1) \left(\overline{xyz} \ \overline{w} + \overline{xyw} \ \overline{z} + \overline{xzw} \ \overline{y} + \overline{yzw} \ \overline{x}\right) \Big. \nonumber \\ 
			&\nonumber \hspace{-0.4cm} \Big. -  (m-1) \left(\overline{xy} \ \overline{zw} + \overline{xz} \ \overline{yw} + \overline{xw} \ \overline{yz} \right)  \Big. \\
			& \hspace{-0.4cm} \Big.+ 2 m \left(\overline{xy} \ \overline{z} \ \overline{w} + \overline{xz} \ \overline{y} \ \overline{w} + \overline{xw} \ \overline{y} \ \overline{z} \right. \Big. \nonumber \\
			& \hspace{+0.6cm} \Big. \left. + \overline{yz} \ \overline{x} \ \overline{w} + \overline{yw} \ \overline{x} \ \overline{z} + \overline{zw} \ \overline{x} \ \overline{y}\right) \Big. \nonumber \\
			& \hspace{-0.4cm} \Big.- 6m \overline{x} \ \overline{y} \ \overline{z} \ \overline{w} \Big.]. \label{eq:c4_estimator}
		\end{align}
		Their structure is similar to that of the cumulants apart from $m$-dependent prefactors [compare Eqs. (\ref{eq:C2})-(\ref{eq:C4})]. A factorized form of $c_4(x,y,z,w)$ is implemented in our SignalSnap library for faster computation and can be found in \ref{app:factorizedC4}.
		The estimators have the property
		$\langle c_i \rangle = C_i$ for finite $m$ (unbiased estimators) and $c_i \rightarrow C_i$ for $m \rightarrow \infty$ (consistency)  \cite{schefczikARXIV2019}. In contrast, naive or natural estimators introduce biases of order $1/m$, which can lead to misleading spectral features—especially at higher orders. An especially striking example will be presented in Section~\ref{sec:UnbiasedEstimation}. 
		
		 It is known that the variance of the estimators decreases towards a constant level with $m$ (see “scaled variance” in Fig. 1 of \cite{schefczikARXIV2019}). This is the reason why SignalSnap has a default value of $m=10$ and not a lower one, although a low $m$ is beneficial for suppressing quasi-correlations for slightly non-stationary signals (see Sec. \ref{sec:QuasiSpectra}). Moreover, the noise (variance) of estimators 
		generally increases with their order \cite{schefczikARXIV2019}.
		
		The use of k-statistics for estimating polyspectra implies the assumption that the vector of Fourier-coefficients $a_k$ is an i.i.d. stochastic variable. This requirement is met approximately by processes $z(t)$ that lose their memory of the past within a time interval that is shorter than window length $T$, where $T$ appears in the calculation of Fourier coefficients. Fourier coefficients of subsequent windows are then in a very good approximation independent from each other. As $z(t)$ is also required to be a stationary process the Fourier coefficients are also identically distributed fulfilling the i.i.d. assumption.  
		We emphasize that polyspectra of processes that exhibit sharp spectral structures may in general not be estimated using windows that are too short for resolving sharp spectral features. Subsequently calculated Fourier coefficients at the frequency of the sharp structure would have a fixed phase relation and, therefore, violate the assumption of independence. 
		
		Last, we remark that the estimation of spectra may in the future benefit from advances in the understanding of estimators. The Fourier-coefficients for finite frequencies (escaping the leakage of the zero-frequency contribution of the signal caused by windowing) are for a stationary process average free. This allows for finding estimators different from the k-statistics.
		In the case of a stochastic variable $x$, it has long been known that $C_3(x,x,x)$ can for $\langle x \rangle = 0$ be estimated via $\overline{x^3}$. It turns out, however, that 
		$m^2[(m-1)(m-2)]^{-1}[\overline{x^3} - 3\overline{x^2} \overline{x} +  \overline{x}^3 ]$ shows in case of a Gaussian dominated process $x$ less noise (this "astonishing" find is mentioned in \cite{anscombe1961}, page 14), while $c_3^{\rm (opt)} = \overline{x^3} - \frac{3(m-1)}{m+1}\overline{x^2} \overline{x}$ is optimal  \cite{schefczikARXIV2019}. A recent advance in estimation in the case of known distributions was reported by Chan \cite{chanJSCS2020}. New versions of SignalSnap may therefore regard the nature of a stochastic signal and adapt its estimators correspondingly.

		\subsection{Error Estimation of Spectral Values}
		Spectral values that are estimated from a finite amount of experimental data can only approximate the ideal values that would follow from an infinite amount of data. SignalSnap calculates a spectral value $S$ by averaging spectral estimates $S_i$  for $N_{\rm p}$ parts of the data with equal size. 
		Considering ${\rm Re}(S_i)$ and ${\rm Im}(S_i)$ as stochastic variables we estimate their variances via 
		the unbiased estimator 
		\begin{equation}
			{\rm Var}\left(x\right) \approx
			\frac{N_{\rm p}}{N_{\rm p} -1} \left( \overline{x^2} - \overline{ x }^2 \right),
		\end{equation}  
		where $x$ corresponds to ${\rm Re}(S_i)$ or ${\rm Im}(S_i)$, respectively.
		The standard error of the average $\overline{x}$ is
		\begin{equation}
			\sigma_{\overline{x}} = \sqrt{\frac{\text{Var}(x)}{N_{\rm p}}}.
		\end{equation}
		which is the basis for visualizing errors in the plot functions of SignalSnap.

\subsection{Current Limitations of Existing Implementations}

Despite the established theoretical framework introduced by Brillinger, practical implementations of cumulant-based higher-order spectra remained incomplete or incorrect. A detailed examination of existing popular libraries highlights two fundamental issues: the absence of proper cumulant-based formulations and the use of biased estimators.

The widely cited MATLAB Higher Order Spectral Analysis (HOSA) toolbox, for example, does not implement Brillinger's cumulant-based trispectrum at all \cite{swami2025hosa}. Instead, it provides functionality only for the so-called Wigner trispectrum, a fundamentally different quantity. According to the HOSA documentation, the Wigner trispectrum is defined using the fourth-order product
\begin{equation}
	r_4(t, \tau_1, \tau_2, \tau_3) = x^*(t - \tau)  x(t - \tau + \tau_1)  x(t - \tau + \tau_2)  x^*(t - \tau + \tau_3),
\end{equation}
where $\tau := (\tau_1 + \tau_2 + \tau_3)/4$. The Wigner trispectrum is then obtained as a three-dimensional Fourier transform of this product,
\begin{align}
	W(t, \omega_1, \omega_2, \omega_3) & \nonumber \\
	& \hspace{-2cm}= \iiint e^{-j (\omega_1 \tau_1 + \omega_2 \tau_2 - \omega_3 \tau_3)} r_4(t, \tau_1, \tau_2, \tau_3) , d\tau_1 d\tau_2 d\tau_3.
\end{align}
Within the library this integral is evaluated only for $\omega_1= \omega_2=- \omega_3$. As such, the Wigner trispectrum implemented in HOSA is neither cumulant-based nor related to Brillinger's definition of the trispectrum and evaluated only partially.

Similarly, the Python-based Higher Order Spectral Analysis toolkit (pyHOSA, hereafter) explicitly claims to implement cumulant-based trispectral analysis \cite{chatterjee2025pyhosa}. However, upon close inspection, it becomes apparent that the implemented estimator is only the incomplete cumulant
\begin{align}
	c_{4,\text{pyHOSA}}(x, y, z, w)  = (\overline{x-\bar{x})(y-\bar{y})(z-\bar{z})(w-\bar{w})},	\label{eq:C4pyHOSA}			
\end{align}
where the second-order cross terms are missing [comp. Eq.~\eqref{eq:factorizedC4}]. This expression is then directly used to calculate their trispectrum. Moreover, even if the cumulant structure had been correctly implemented, the omission of finite-sample correction prefactors ($m$-dependent factors from the $k$-statistics) would still result in a biased estimator. Additionally, a window function dependent normalization is missing making the spectral values dependent on the window function and window length. As a result, spectra estimated by pyHOSA, may show false offsets and artifacts unrelated to true signal correlations.

		\begin{figure}[t]
	\centering
	\includegraphics[width=\columnwidth]{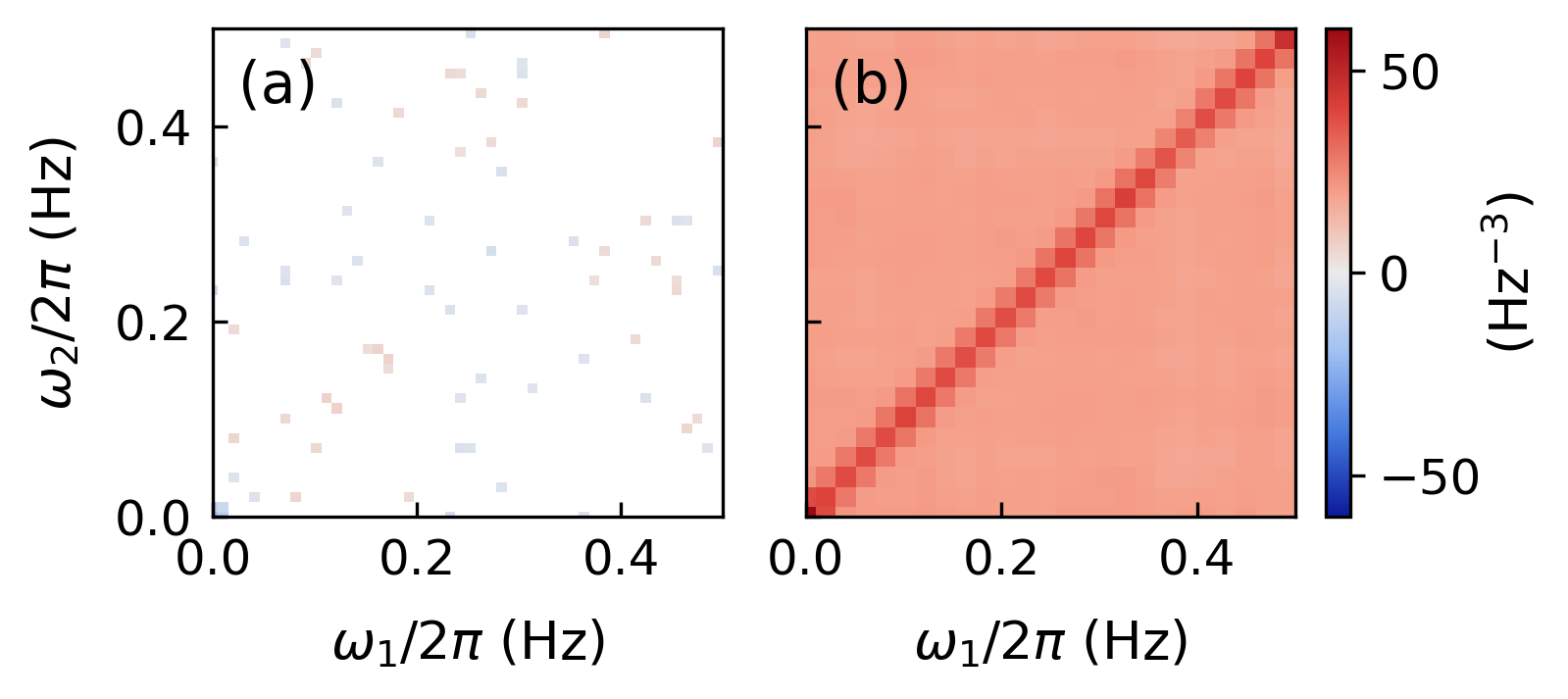}
	\caption{Comparison of trispectra estimated for white Gaussian noise (length $N = 10^5$ with zero mean and unit variance) using two different methods. 
		(a) Trispectrum computed with SignalSnap’s unbiased cumulant-based estimator. As expected, the spectrum is statistically consistent with zero, showing no structure beyond noise level. 
		(b) Trispectrum computed using the moment-based estimator found in the pyHOSA library. The result incorrectly shows strong non-zero values, especially along the diagonal, illustrating the spurious spectral structures caused by incorrect estimator choice. Both panels share identical axes and color scale.}
	\label{fig:hosa_comparison}   
\end{figure}

To demonstrate these shortcomings, we generated a white Gaussian noise signal of length $N = 10^5$ with zero mean and unit variance, which should results in a statistical zero for the higher-order spectra. The signal was analyzed using both the pyHOSA estimator and the SignalSnap cumulant-based estimator. The results are compared in Figure~\ref{fig:hosa_comparison}. Panel (a) shows the trispectrum computed using SignalSnap’s unbiased cumulant-based estimator, which remains statistically consistent with zero. Panel (b) shows the trispectrum obtained using the estimator from pyHOSA, which incorrectly displays a pronounced offset and a diagonal structure. This is due to the fact that missing second-order terms in Eq.~\eqref{eq:C4pyHOSA} are non-zero, since the mean of squared Gaussian is positive. These results clearly illustrate the necessity of employing unbiased, cumulant-based estimators to ensure meaningful and artifact-free higher-order spectral analysis.

		\section{Signals and their Polyspectra}
		\label{sec:first_example}
		
		To develop an intuitive understanding of higher-order spectra, concrete examples are essential. Unfortunately, such examples are scarce in the literature. Foundational works such as those by Mendel~\cite{mendelIEEE1991} and Nikias~\cite{NikiasIEEE1993} provide theoretical insight and application-oriented discussions, but rarely include plots of bispectra or trispectra for illustrative signals. To address this gap, we present here four instructive examples that highlight how nonlinear signal operations affect higher-order spectral features.

		Each example is based on the time-dependent position \( x_i(t) \) of two uncoupled stochastically driven harmonic oscillators. Using the velocities \( v_i(t) = \frac{dx_i(t)}{dt} \), the stochastic equations of motion are given by
		\begin{align}
			dv_i &= - 2 \gamma v_i \, dt +  \omega_i^2 x_i \, dt + \sigma\,dW_i \nonumber \\
			dx_i &= v_i \, dt\,, \label{eq:linear_filter}
		\end{align}
		where \( dW_i \) is the increment of a Wiener process with  $\Gamma_i(t) =dW_i/dt$ being $\delta$-correlated white noise. We solve these equations numerically over a total duration of \(T = 10^3\,\si{s}\), using frequencies \(\omega_1/2\pi = \SI{2}{kHz}\), \(\omega_2/2\pi = \SI{3}{kHz}\), damping \(\gamma = \SI{1}{kHz}\), and noise strength \(\sigma = \SI{1}{kHz^{3/2}}\).
		
\begin{figure*}[tbhp]
			\centering
			\includegraphics[width=\textwidth]{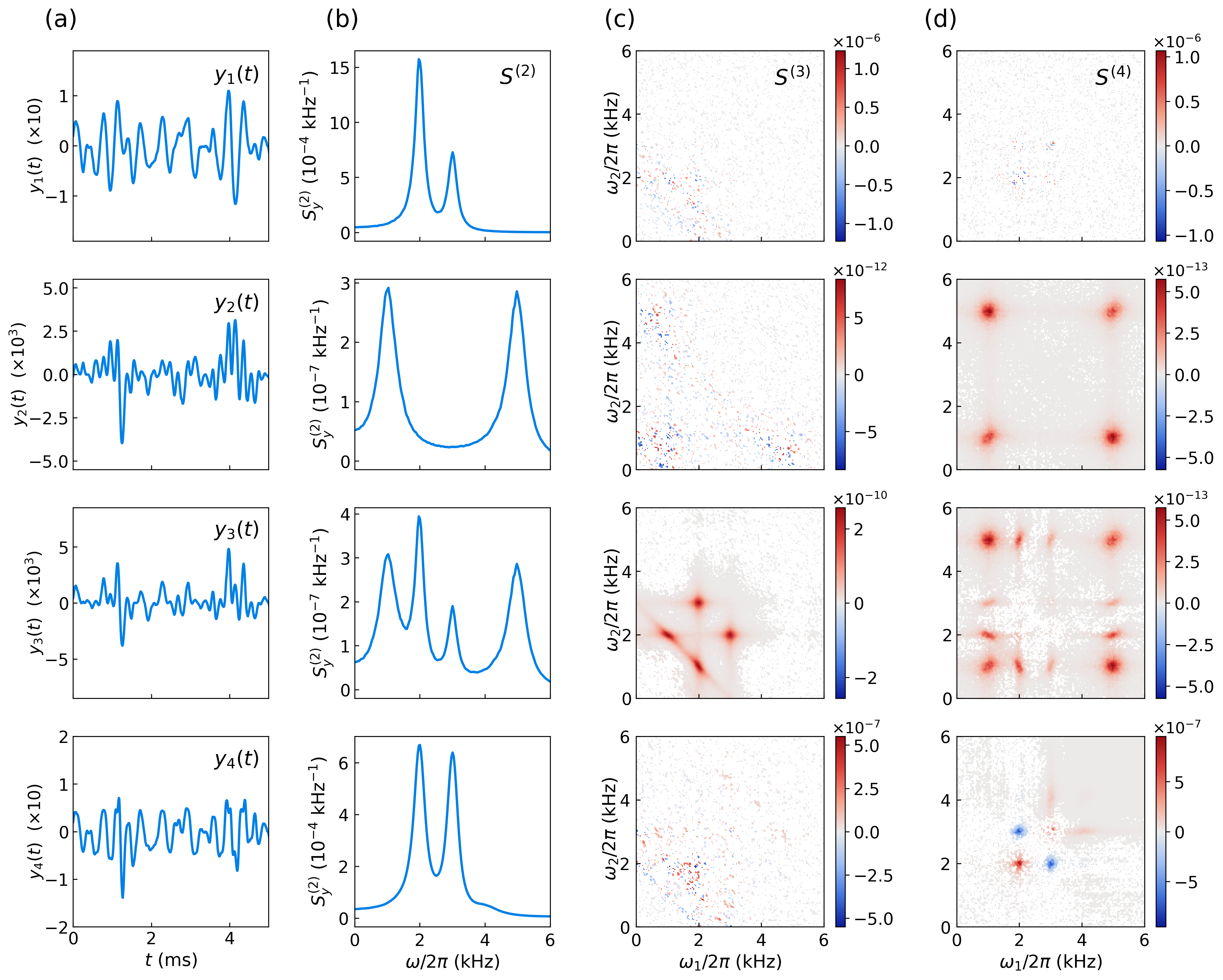}
			\caption{Comparison of signals $y_1(t)$ to $y_4(t)$ and their  spectra $S^{(2)}$, $S^{(3)}$, $S^{(4)}$. The signals $y_j(t)$ are constructed from Gaussian signals $x_1(t)$ and $x_2(t)$ via linear and nonlinear operations.    \textit{Row 1:} The signal of the linear superposition $y_1(t)=x_1(t)+x_2(t)$ shows Lorentzian peaks at \SI{2}{kHz} and \SI{3}{kHz} while the higher-order spectra show no significant contributions.
				\textit{Row 2:} The signal $y_2(t)=x_1(t)x_2(t)$ from multiplicative mixing shows in $S^{(2)}$ frequency components only at the sum and difference frequencies \SI{5}{kHz} and \SI{1}{kHz}, respectively. The bispectrum remains insignificant, while the trispectrum reveals positive correlations among the new frequency components. 
				\textit{Row 3:} The signal $y_3(t)=x_1(t) x_2(t)+a x_1(t) + a x_3(t)$ with $a=0.015$ shows four peaks, namely at the frequencies of $x_1(t)$, $x_2(t)$ and at their difference and sum frequencies. The bispectrum shows significant contributions due to the phase-locked mixing of frequencies. The trispectrum shows positive correlations between linear and mixed contributions to $y_3(t)$.
				\textit{Row 4:} The power spectrum $S^{(2)}$ of signal $y_4(t)=x_1(t)(1-bx_2(t)^2)+x_2(t)$ reveals strong peaks at the base frequencies  \SI{2}{kHz} and \SI{3}{kHz} and a weaker additional contribution at \SI{4}{kHz}. The bispectrum shows no significant contributions. The trispectrum reveals negative correlations between the base frequencies that were caused by the third-order non-linear mixing of $x_1(t)$ and $x_2(t)$. The figure demonstrates that higher spectra can be viewed as fingerprints of important relations in a stochastic process. Spectra $S_z^{(3)}$ and $S_z^{(4)}$ are given in units of kHz$^{-2}$ and kHz$^{-3}$, respectively.}
			\label{example_overview}   
		\end{figure*}

		Figure~\ref{example_overview} summarizes the power spectrum \(S^{(2)}\), bispectrum \(S^{(3)}\), and the trispectrum \(S^{(4)}\) for each constructed signal. Spectral values within the estimated $3\sigma$ noise floor are rendered in white.
		
		\paragraph{Linear superposition}
		The first row of Figure~\ref{example_overview} shows the polyspectra of the sum
		\begin{align}
			y_1(t) = x_1(t) + x_2(t)
		\end{align}
		of the oscillator positions. The oscillator with the higher frequency is governed by a stronger spring that drives the mass towards the zero position. Consequently, it exhibits smaller amplitudes compared to the oscillator which oscillates at a lower frequency. The power spectrum thus exhibits a large peak at low frequency and a small peak at the higher frequency.  The higher-order spectra $S^{(3)}$ and $S^{(4)}$  of $y_1(t)$ are expected to be zero and exhibit only a few points outside the $3\sigma$ error bound. This absence of correlations in the higher-order spectra is a consequence of the fact that  $x_1(t)$ and $x_2(t)$ follow from white Gaussian noise and linear filtering. Gaussian noise is known to have vanishing higher-order correlations \cite{gardinerBOOK2009}. 
		
		\paragraph{Multiplicative mixing}
		The second row of Figure~\ref{example_overview} shows the effect of multiplicative mixing of $x_1$ and $x_2$ resulting in a new signal 
		\begin{align}
			y_2(t) = x_1(t) x_2(t).
		\end{align}
		The power-spectrum $S^{(2)}$ exhibits peaks at the sum (\SI{5}{kHz}) and difference (\SI{1}{kHz}) of the frequencies $\omega_1$ and $\omega_2$ of $x_1(t)$ and $x_2(t)$. The peaks in the “mixed” spectrum are broader than those of the initial spectrum as the widths of the initial spectral peaks get convolved in the mixed signal.

		The bispectrum $S^{(3)}$ of $y_2(t)$ shows no significant non-zero contributions. Contributions to the bispectrum arise from the cumulant $C_3(z(\omega_1), z(\omega_2), z^*(\omega_1 + \omega_2))$. The signal $y_2$ has contributions at $1$~kHz and $5$~kHz but none at $6$~kHz. Consequently, no contribution to $S^{(3)}(\omega_1,\omega_2)$ is expected at that pair of frequencies. The same holds for the pairs  $1$~kHz,  $1$~kHz and $5$~kHz, $5$~kHz.
		The cut of the trispectrum $S^{(4)}$, however, displays significant positive diagonal and off-diagonal contributions. The cumulant $C_4(z(\omega_1),z^*(\omega_1), z(\omega_2),z^*(\omega_2))$ governs $S_z^{(4)}$ and is independent on the 
		phase relations between $z(\omega_1)$ and  $z(\omega_2)$. The positive spectral contribution of the frequency pair $1$~kHz and $1$~kHz reveals
		that signal $y_2$ at frequency $1$~kHz is more noisy than Gaussian noise. 
		This followed from the multiplication of two initial Gaussian noises at frequencies 
		$2$~kHz and $3$~kHz. Similarly,  $S^{(4)}$ reveals positive correlations at
		the frequency pairs $1$~kHz and $5$~kHz, and $5$~kHz and $5$~kHz.

		\paragraph{Mixed signal with linear components}
		Figure~\ref{example_overview}  shows in the third row the polyspectra of 
		\begin{align}
			y_3(t) = x_1(t)x_2(t) +ax_1(t) + ax_2(t) 
		\end{align}
		with $a=0.015$. 
		The linear terms $ax_1(t)$ and $ax_2(t)$ ensure that the signal keeps contributions with frequencies at $\omega_1 = 2$~kHz and $\omega_2 = 3$~kHz that now appear 
		in the power spectrum $S^{(2)}$. Contributions at $1$~kHz and $5$~kHz appear due to the mixing of $x_1$ and $x_2$ by the first term. This causes non-negligible contributions to the bispectrum $S^{(3)}$ as now the contributions at the three frequencies $2$~kHz, $3$~kHz, and $5$~kHz are correlated. Similarly, contributions at $1$~kHz, $2$~kHz, and $3$~kHz are correlated. Note, that a term $- x_1(t) x_2(t)$ in $y_3$ instead of $x_1(t) x_2(t)$ would have led to a bispectrum with a negative sign.  
		The fourth-order spectrum $S^{(4)}$ shows positive correlations between all signal contributions that appear as peaks in $S^{(2)}$ except for the frequency pairs $2$~kHz, $2$~kHz and $3$~kHz, $3$~kHz. These frequency contributions are purely Gaussian and therefore exhibit no higher-order correlations.

		\paragraph{Amplitude modulation and suppression}
		As a last example, we discuss polyspectra [Figure \ref{example_overview}, last row] of the signal 
		\begin{align}
			y_4(t) = x_1(t)(1- b x_2(t)^2) + x_2(t),
		\end{align}
		where  $b = 500$.  The term $ b x_2(t)^2$ has a strong contribution at zero frequency and therefore leads in the first term to a stochastic modulation of the intensity of $x_1(t)$.
		Whenever the intensity of $x_2$ at $3$~kHz is high, the intensity of the $x_1(t)(1- b x_2(t)^2)$ contribution is lowered at frequency $2$~kHz. This anti-correlation is revealed in the negative peaks appearing in the spectrum $S^{(4)}$ at the frequency pairs $2$~kHz and $3$~kHz.  The positive peak at $2$~kHz, $2$~kHz is a consequence of the stochastic modulation of the initial $x_1$ signal which results in a signal at $2$~kHz which is noisier than usual Gaussian noise. The conditions for bispectral components are not fulfilled leading to no significant contributions to $S^{(3)}$.

The examples above illustrate how both linear and nonlinear signal operations manifest as distinctive features in higher-order spectra. Polyspectra thus provide a powerful window into structures and dependencies present in stochastic processes.

\section{Importance of Unbiased Estimation}
\label{sec:UnbiasedEstimation}

Up to this point, we have repeatedly emphasized the importance of using unbiased estimators when calculating higher-order spectra. In this section, we demonstrate the tangible consequences of using biased versus unbiased estimators. The difference is not just a slight shift in amplitude. Biased estimation can introduce entirely artificial spectral features in the fourth-order spectrum. 

				\begin{figure*}[t]
	\centering
	\includegraphics[width=\textwidth]{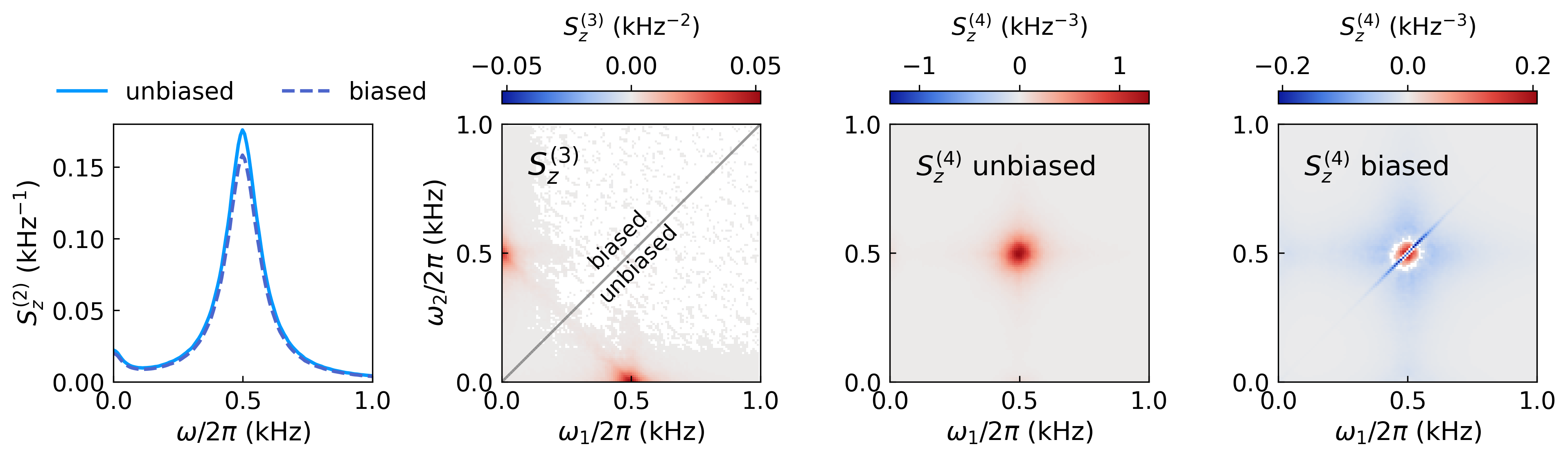}
	\caption{Polyspectra of a non-Gaussian signal $z(t)$ [see Eq. (\ref{eq:signalz})]. Spectra from unbiased and biased estimators are compared. The unbiased spectra $S_z^{(2)}$ and $S_z^{(3)}$ are scaled down by a constant prefactor $m/(m-1)$ and $m^2/[(m-1)(m-2)]$. The biased spectrum $ S_z^{(4)}$ exhibits false structures in the spectrum that are absent from the unbiased estimate. Clearly, the use of unbiased estimators is not an option for calculating fourth-order polyspectra (albeit we are not aware of any unbiased estimates in the previous literature). }
	\label{fig:unbiased_plot}   
\end{figure*}

Our method for estimating higher-order spectra uses exclusively the multi-variate version of the k-statistics to calculate cumulant estimates of the Fourier coefficients of the signal. The k-statistics exhibit factors that depend on the number $m$ of samples. Those prefactors assume unity in the limit $m \rightarrow \infty$ but are vital at finite $m$ for yielding unbiased estimates.  In contrast, the previous literature on the estimation of polyspectra relies on estimators that are at best asymptotically unbiased \cite{trappMSSP2021}. We compare in Fig.~\ref{fig:unbiased_plot} polyspectra of the same signal calculated for $m =10$. One batch of spectra is estimated with the k-statistics, the other one is estimated with the so-called natural estimator where all prefactors are unity. 
The signal 
\begin{equation}
	z(t) = \left(z_1(t) + 0.1\right)\,\left(z_2(t) + 0.1\right) \label{eq:signalz}
\end{equation}
is generated from the mixing of two gaussian signals. The first signal, \(z_1(t)\), is low-pass-filtered white noise obtained via the transfer function in Eq.~\eqref{eq:intro_model} with parameters \(S_0 = 25/\pi\) and \(\gamma/2\pi = \SI{0.4}{kHz}\). The second signal, \(z_2(t)\), is produced by filtering white noise through a Lorentzian bandpass filter, whose realization is described in \ref{app:bandpass}. For this filter, the parameters are set to \(\omega_1/2\pi = \SI{0.5}{kHz}\) and \(\gamma_1/2\pi = \SI{0.04}{kHz}\).
The spectra $S^{(2)}$ and $S^{(3)}$ exhibit only a slight rescaling caused by the missing prefactors $m/(m-1)$ and $m^2/[(m-1)(m-2)]$, respectively. The spectrum $S^{(4)}$ calculated from the biased (natural) estimator, however, exhibits strong false features. The prefactors $m+1$ and $m-1$ for different contributions to $c_4$ in the square bracket of the factorized version of $c_4$ no longer allow for relating the natural estimator and the k-statistics by a common prefactor [see Eq. (\ref{eq:factorizedC4})]. 
Clearly, the use of unbiased estimators can lead to wrong estimates of spectra resulting in false additional features. 
An accurate, unbiased estimate of experimental spectra was essential in a recent work involving the comparison of experimental to theoretical spectra. The presence of artifacts, such as those in Fig. \ref{fig:unbiased_plot}, would have compromised the validity of the fitting process \cite{sifftPRR2021,SifftPRA2024}.

		\section{Polyspectra of multiple channels}
		\label{sec:MultChann}

Until now, our analysis has centered on single-channel signals, where all spectral information originates from a single time-dependent stochastic process. However, many experiments involve multiple sensors recording data simultaneously. These multichannel signals often exhibit subtle interdependencies - whether between distinct physical observables or spatially separated locations - that are invisible to single-channel approaches.
		
Multi-channel polyspectra provide a powerful framework to quantify such inter-signal correlations. For example, cross-polyspectra can be used to detect correlations between a frequency-modulated signal and encoded amplitude, or to uncover nonlinear couplings between the $x$ and $y$ coordinates in the motion of a driven rotator. They are also essential for multi-detector setups, where one seeks to distinguish true signal correlations from independent noise across channels. 
		
		SignalSnap implements the following multi-channel polyspectra 
		\begin{align}
			C_n(z_1(&\omega_1), \dots, z_n(\omega_n)) = \nonumber \\
			&2\pi \delta(\omega_1+\dots+\omega_n)S_{z_1, \dots, z_n}^{(n)}(\omega_1,\dots,\omega_{n-1})\,	  \label{eq:defPolyspectraMulti}
		\end{align}
		as a generalization of Brillinger's single-channel spectra Eq.~(\ref{eq:defPolyspectra}). We note that Mendel gave a generalization 
		of polyspectra to a vector of stochastic processes, where his definition regarded all possible combinations of different channels \cite{mendelIEEE1991}.
		The SignalSnap implementation of Eq. (\ref{eq:defPolyspectraMulti}) allows the user to specify the combination of channels for calculating a polyspectrum.  
		
		The single-channel equations \eqref{eq:c2_estimator}--\eqref{eq:c4_estimator} for calculating poly\-spectra are easily generalized to the multi-channel case.
		The polyspectra for signals $z_1(t), \dots, z_n(t)$ are 
		\begin{align}
			S^{(2)}_{z_1, z_2}(\omega_k) &\approx \frac{N C_2(a_k, b_k^*)}{T \sum_{i = 0}^{N-1} g_i g^*_i} \label{eq:MultiS2}\\
			S^{(3)}_{z_1, z_2, z_3}(\omega_k, \omega_l) &\approx \frac{N C_3(a_k, b_l, c_{k+l}^*)}{T \sum_{i = 0}^{N-1} g_i^2 g^*_i} \\
			S^{(4)}_{z_1, z_2, z_3, z_4}(\omega_k, \omega_l,\omega_p) &\approx \frac{N C_4(a_k, b_l, c_p, d_{k+l+p}^*)}{T \sum_{i = 0}^{N-1} g_i^3 g^*_i}.
		\end{align}
		The Fourier coefficients $a_k, b_k, \dots$ correspond to signals $z_1, z_2, \dots$, respectively. Please note, that the single-channel case is recovered for identical signals $z_1(t) \equiv z_2(t) \equiv \dots$ and $a_k \equiv b_k \equiv \dots$.

		SignalSnap implements only a two-dimensional cut through the fourth-order spectrum
		\begin{equation}
			S^{(4)}_{z_1, z_2, z_3, z_4}(\omega_k, \omega_l) \approx \frac{N C_4(a_k, b_k^*, c_l, d_l^*)}{T \sum_{i = 0}^{N-1} g_i^2 (g^*_i)^2}.
		\end{equation}
		
		The multi-channel polyspectra exhibit fewer symmetries and fewer general properties than their single-channel counterparts. While $S^{(2)}_z(\omega)$ is always non-negative and symmetric (i.e. $S^{(2)}_z(-\omega) = S^{(2)}_z(\omega) \geq 0$ ), the cross-correlation spectrum $S^{(2)}_{xy}$ is in general complex where the relations $S^{(2)}_{xy}(-\omega) = [S^{(2)}_{xy}(\omega)]^* = S^{(2)}_{yx}(\omega)$ hold. The decreasing number of symmetries with respect to the single-channel case increases the computational cost for calculating higher-order polyspectra in the multi-channel case.

		\subsection{Example: Random telegraph noise switches frequency of harmonic oscillator}
\begin{figure*}[t]
			\centering
			\includegraphics[width=\textwidth]{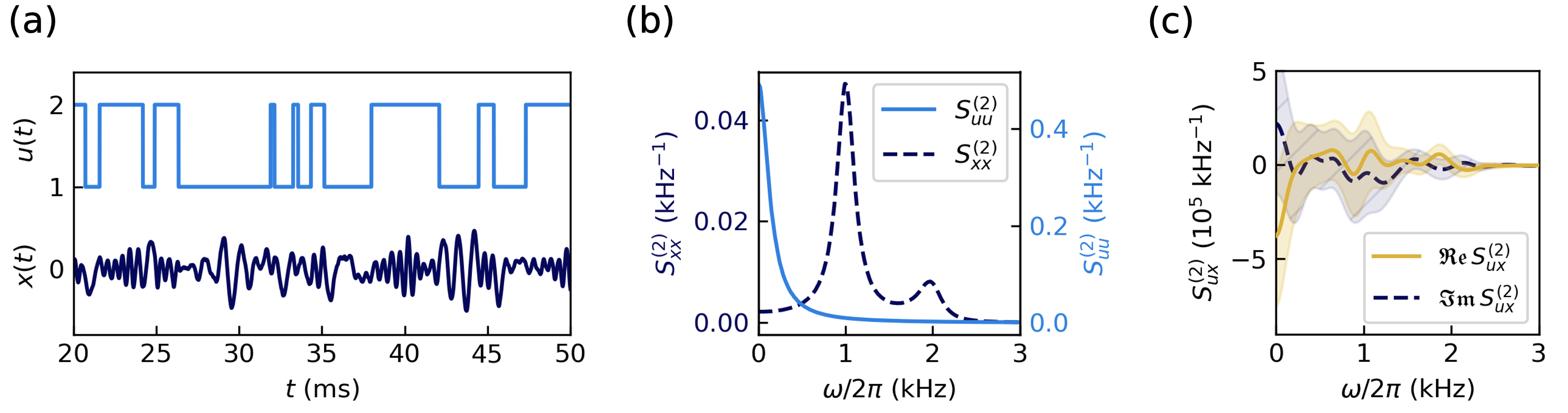}
			\caption{(a) Samples of signals $u(t)$ and $x(t)$, where $u(t)$ is random telegraph noise. The signal $x(t)$ represents the position of a stochastically driven harmonic oscillators whose frequency switches between two frequencies as $u(t)$ changes values. (b) The power-spectra $S^{(2)}_{uu}$ and $S^{(2)}_{xx}$ of the two signals. (c) The cross-correlation spectrum $S^{(2)}_{ux}$ shows no significant contributions in both, its real and imaginary part. The $3\sigma$-error bounds are indicated as shaded areas. 
				The higher-order two-channel polyspectra of $u(t)$ and $x(t)$ are shown in Fig. \ref{fig:multi_det_S3_S4}. }     
			\label{fig:multi_det_S2}
		\end{figure*}

		To illustrate the capabilities of multi-channel polyspectra, we now present a simple, yet informative example that can be well analyzed and interpreted. This example is designed to showcase two key features: non-Gaussian inter-channel couplings and a broken time-reversal symmetry.
		We consider a system composed of two correlated signals. The signal $u(t)$ is random telegraph noise that switches between the values 1 and 2 with the transitions rates $\lambda_{12}=0.3\,\text{ms}^{-1}$ and $\lambda_{21}=0.6\,\text{ms}^{-1}$ [see upper line in Fig. \ref{fig:multi_det_S2}(a)].
		The signal $x(t)$ represents the time-dependent elongation of a stochastically driven harmonic oscillator whose frequency switches with $u(t)$ between two values given by $\omega_0 u(t)$ with $\omega_0 = 1$~kHz. The equation of motion is 
		\begin{align}
			dv &= - 2 \gamma v \,dt +  (\omega_0 u(t))^2 x \,dt + \sigma\,dW\,,\\
			dx &= v \,dt\,,
		\end{align}
		where $\sigma=2\,\text{kHz}^{3/2}$ and $\gamma=0.5\,\text{kHz}$.

		The polyspectra of Figure \ref{fig:multi_det_S2} were calculated with SignalSnap from $u(t)$ and $x(t)$ for a time interval of $4 \times10^3$~s
		with a temporal resolution of $10^{-6}$~s corresponding to $4\times10^9$ pairs of datapoints. The distance between adjacent points in the spectra is 10~Hz. The computation time for a two-channel fourth-order spectrum in Figure \ref{fig:multi_det_S3_S4} was about 150~s on a PC with a Nvidia RTX 4090 GPU.
		Figure \ref{fig:multi_det_S2} (a) displays the behavior of $u(t)$ and $x(t)$ in a $30$~ms time interval. The dependency of the oscillator frequency on the modulation signal $u(t)$ is clearly visible. Fig.~\ref{fig:multi_det_S2} (b) shows the power spectrum $S^{(2)}_{xx}$ of the oscillator which reveals a peak at $\SI{1}{kHz}$ and another peak at $\SI{2}{kHz}$ as expected. The power spectrum of the modulation signal $S_{uu}^{(2)}$ shows a characteristic Lorentz-shaped spectrum centered at zero frequency as expected for two state telegraph noise (compare  Fig.~\ref{fig:intro_plot}) \cite{sifftPRR2021}. Since the spectral overlap between the spectra of $u(t)$ and $x(t)$ is very small, the cross-correlation spectrum $S^{(2)}_{ux}$ reveals no significant contributions [Fig.~\ref{fig:multi_det_S2} (c)].
		
		\begin{figure*}[t]
			\centering
			\includegraphics[width=\textwidth]{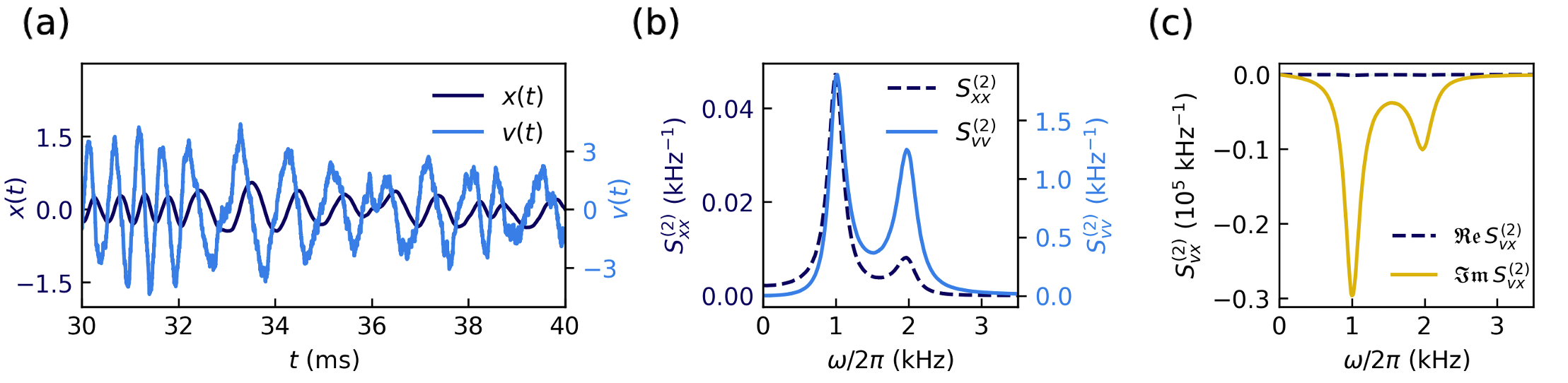}
			\caption{(a) Position $x(t)$ and velocity $v(t)$ of the process discussed in the text. (b) Power-spectra of $x(t)$ (dashed line) and $v(t)$ (solid line). (c) The cross-correlation spectrum $S_{vx}^{(2)}$ has a zero real part and a negtive maginary part with peaks at $\SI{1}{kHz}$ and $\SI{2}{kHz}$ showing that $x(t)$ is behind $v(t)$ in time for both frequencies.}
			\label{fig:multi_det_S2xv}
		\end{figure*}
		
		The velocity $v(t) = dx(t)/dt$ and $x(t)$ show a strong negative imaginary part in the cross-correlation spectrum $S^{(2)}_{vx}(\omega)$ [see Figure \ref{fig:multi_det_S2xv}(c)]. This comes as no surprise as strictly $v(\omega) = - j \omega x(\omega)$ and therefore $S^{(2)}_{vx}(\omega) = -j \omega S^{(2)}_{xx}(\omega)$ [cmp. Eq. (\ref{eq:C2S2}]. This implies that the stochastic vector $[x(t), v(t)]$ shows no time-inversion symmetry (see \ref{app:timeinversion}).

		We like to stress that, in general, a violation of time-inversion symmetry by a stochastic vector does not imply that the individual components violate time-inversion symmetry.
		A particularly transparent example is the two-dimensional process
		$[\Gamma_1(t), \Gamma_2(t)]$, where $\Gamma_1(t)$ is $\delta$-correlated white noise and $\Gamma_2(t) = \Gamma_1(t + \Delta t)$ is a time-shifted copy of $\Gamma_1(t)$.
		Both components are invariant under time reversal. Their joint statistics, however, is not: The second-order cross-spectrum
		$S_{\Gamma_1,\Gamma_2}^{(2)}(\omega) = e^{j \omega \Delta t}$ follows from $\left\langle\Gamma_1(t) \Gamma_2(t+\tau)\right\rangle= \delta(\tau-\Delta t)$ via Eq. \eqref{S2_autocorr} and exhibits imaginary contributions which imply a violation of time-reversal symmetry.

		\begin{figure*}
			\centering
			\includegraphics[width=\textwidth]{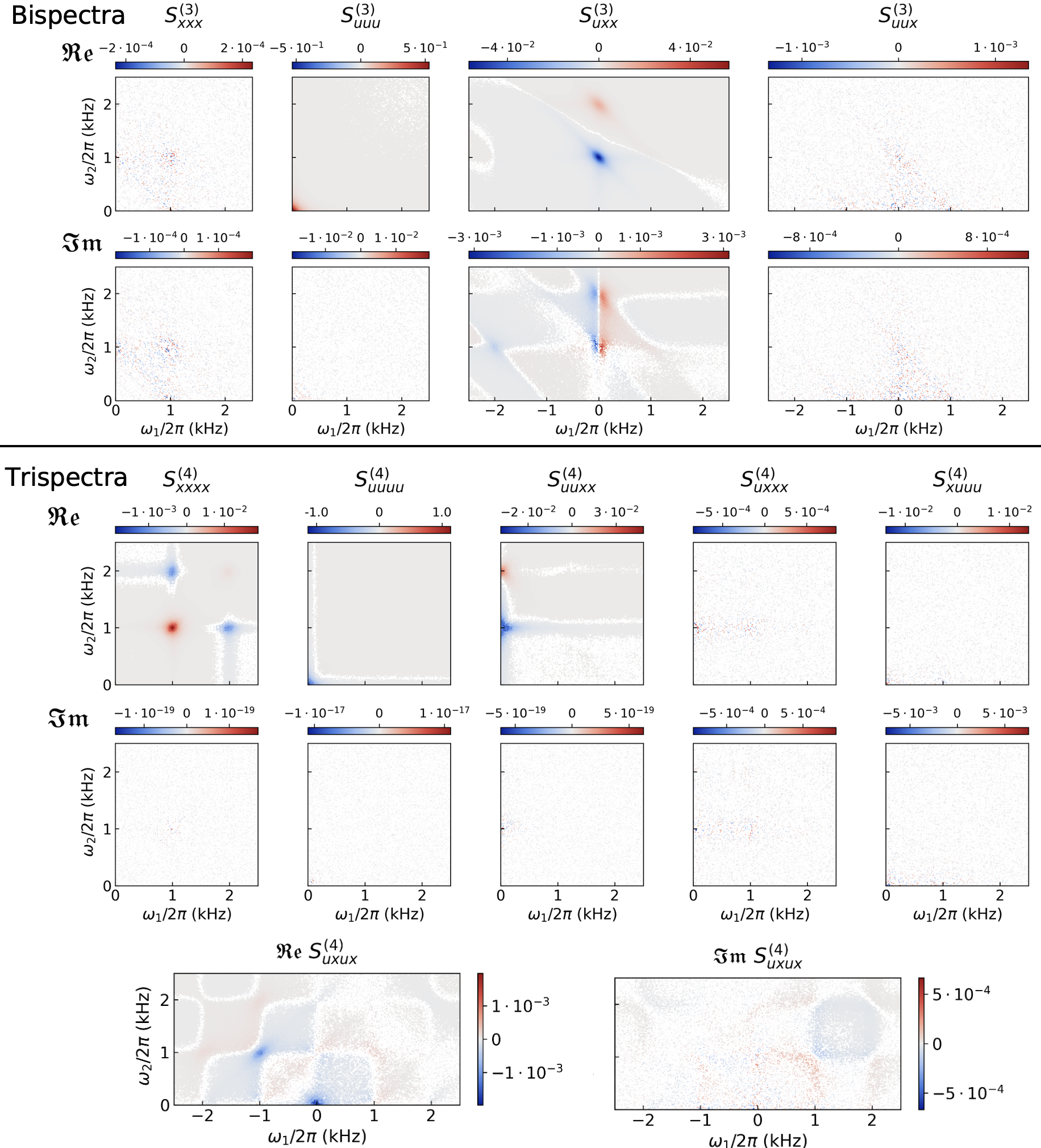}
			\caption{Two-channel bispectra $S^{(3)}$ and trispectra $S^{(4)}$ of the signals $u(t)$ and $x(t)$ (see Figure \ref{fig:multi_det_S2}). The specific selection of spectra $S^{(3)}$ and $S^{(4)}$ shown here for different combinations of $u(t)$ and $x(t)$ covers all information on the signals contained in third and fourth-order frequency resolved two-dimensional spectra. Other spectra like e.g. $S^{(4)}_{uxxu}(\omega_1,\omega_2)$ contains the same information as  $S^{(4)}_{uxux}(\omega_1,\omega_2)$ (see Section \ref{sec:symm} on symmetry). The significant imaginary part of $S^{(3)}_{uxx}$ proofs that the stochastic vector $[u(t),x(t)]$ has broken time-inversion symmetry (see \ref{app:timeinversion}). The overall structure of the spectra is explained in Section \ref{sec:MultChann}. Spectra $S_z^{(3)}$ and $S_z^{(4)}$ are given in units of kHz$^{-2}$ and kHz$^{-3}$, respectively.}
			\label{fig:multi_det_S3_S4}
		\end{figure*}

		Next, we discuss and interpret the spectra of higher-order. The upper panel of Figure \ref{fig:multi_det_S3_S4} shows all four third-order spectra $S^{(3)}_{xxx}$, $S^{(3)}_{uuu}$, $S^{(3)}_{uxx}$, and $S^{(3)}_{uux}$ that need to be distinguished for the two signals $u(t)$ and $x(t)$. Due to the high symmetry of the spectra  
		$S^{(3)}_{xxx}$ and $S^{(3)}_{uuu}$ only a region of positive frequencies needs to be displayed (see Sec. \ref{sec:symm}). Due to the lower symmetry of $S^{(3)}_{uxx}$ and $S^{(3)}_{uux}$  only for one axes the positive side is sufficient to be displayed (in our case $\omega_2$ axes).
		The spectrum $S^{(3)}_{xxx}$ of the oscillator dynamics $x(t)$ exhibits no significant phase-sensitive correlations between the frequencies $1$~kHz and $2$~kHz. This may come as a surprise as the phase of $x(t)$ before a jump of $u(t)$ will be correlated with the phase of $x(t)$ after the jump ($x(t)$ and $v(t)$ will, e.g., not change sign during the jump event).  
		However, $C_3(x(\omega_1/2\pi = 1~{\rm kHz}), x(\omega_2/2\pi = 1~{\rm kHz}), x(\omega_3/2 \pi = -2~{\rm kHz}))$ correlates {\it twice} the phase of the signal contribution at $1$~kHz with the phase of the signal contribution at $2$~kHz which leads to a vanishing cumulant and therefore no signature in the spectrum.
		The spectrum $S^{(3)}_{uuu}$ of the random telegraph noise $u(t)$ exhibits the typical non-zero signature behavior around zero frequencies \cite{sifftPRR2021}. 
		The real part of spectrum $S^{(3)}_{uxx}$ exhibits a negative contribution around the position $(0, 1)$~kHz and a positive contribution at $(0, 2)$~kHz.
		The corresponding cumulants $C_3(u(\omega_1/2\pi = 0~{\rm kHz}), x(\omega_2/2\pi = 1~{\rm kHz}), x(\omega_3/2 \pi = -1~{\rm kHz}))$   and
		$C_3(u(\omega_1/2\pi = 0~{\rm kHz}), x(\omega_2/2\pi = 2~{\rm kHz}), x(\omega_3/2 \pi = -2~{\rm kHz}))$ reveal that the intensity of $x(t)$ is correlated with the sign of $u(t)-\langle u(t) \rangle$.  As the frequency $1$~kHz of $x(t)$ appears for $u(t) = 1$ and $\langle u(t) \rangle > 1$ we find a negative contribution to $S^{(3)}_{uxx}$  at $(0,1)$~kHz.
		Similarly, the positive contribution is explained at $(0,2)$~kHz.

		We find a significant imaginary part of $S^{(3)}_{uxx}$ which proves that there is no time-inversion symmetry of the stochastic vector $[u(t),x(t)]$ (see \ref{app:timeinversion}). 
		We emphasize that the imaginary parts of the previously discussed spectra $S^{(3)}_{xxx}$ and $S^{(3)}_{uuu}$ are zero, which would leave open the possibility for preserved time-inversion symmetry of $x(t)$, $u(t)$, and the stochastic vector $[u(t),x(t)]$. 
		However, only a non-zero imaginary part implies absent time-inversion symmetry, but not the other way around.

		The spectrum  $S^{(3)}_{uux}$ exhibits no significant contributions. Since the power spectrum of signal $u(t)$ is centered around zero the spectrum  
		$S^{(3)}_{uux}$ could at best reveal a correlation with the signal $x(t)$ around its own zero-frequency contribution which, however, is extremely weak.

		The trispectrum $S^{(4)}_{xxxx}$ exhibits a clear anti-correlation between the frequency contributions to $x(t)$ at $1$~kHz and $2$~kHz. This is easily explained as the oscillator jumps between two frequencies. The appearance of one frequency in $x(t)$ excludes the appearance of the other frequency leading to a strong anti-correlation.
		The trispectrum $S^{(4)}_{uuuu}$ shows the usual structure for random telegraph noise \cite{sifftPRR2021}.
		The trispectrum $S^{(4)}_{uuxx}$ exhibits a clear anti-correlation for the intensity of $u(t)$ around frequency $0$~kHz with the
		intensity of the $1$~kHz contribution of $x(t)$. The switching rates of the telegraph model tell us that the oscillator is predominantly in the $1$~kHz state. A temporal increase of the intensity of $u(t)$ indicates a larger than usual switching dynamics which will lead to a reduction of the oscillator being in the $1$~kHz state. Consequently, the intensities of $u(t)$ at $0$~kHz and $x(t)$ at $1$~kHz are anti-correlated as found in the spectrum. The reverse holds true for the $2$~kHz contribution to $x(t)$ which exhibits a positive correlation in $S^{(4)}_{uuxx}$ with $u(t)$.
		The spectra $S^{(4)}_{uxxx}$ and $S^{(4)}_{xuuu}$ exhibit no correlations probably because the spectral overlap of $u(t)$ and $x(t)$ is not given.
		
		The real part of $S^{(4)}_{uxux}$ shows a weak negative correlation for which we currently have no interpretation.

		
		\section{Quasi-polyspectra of non-stationary signals}
		\label{sec:QuasiSpectra}
		\begin{figure*}[tb]
			\centering
			\includegraphics[width=\textwidth]{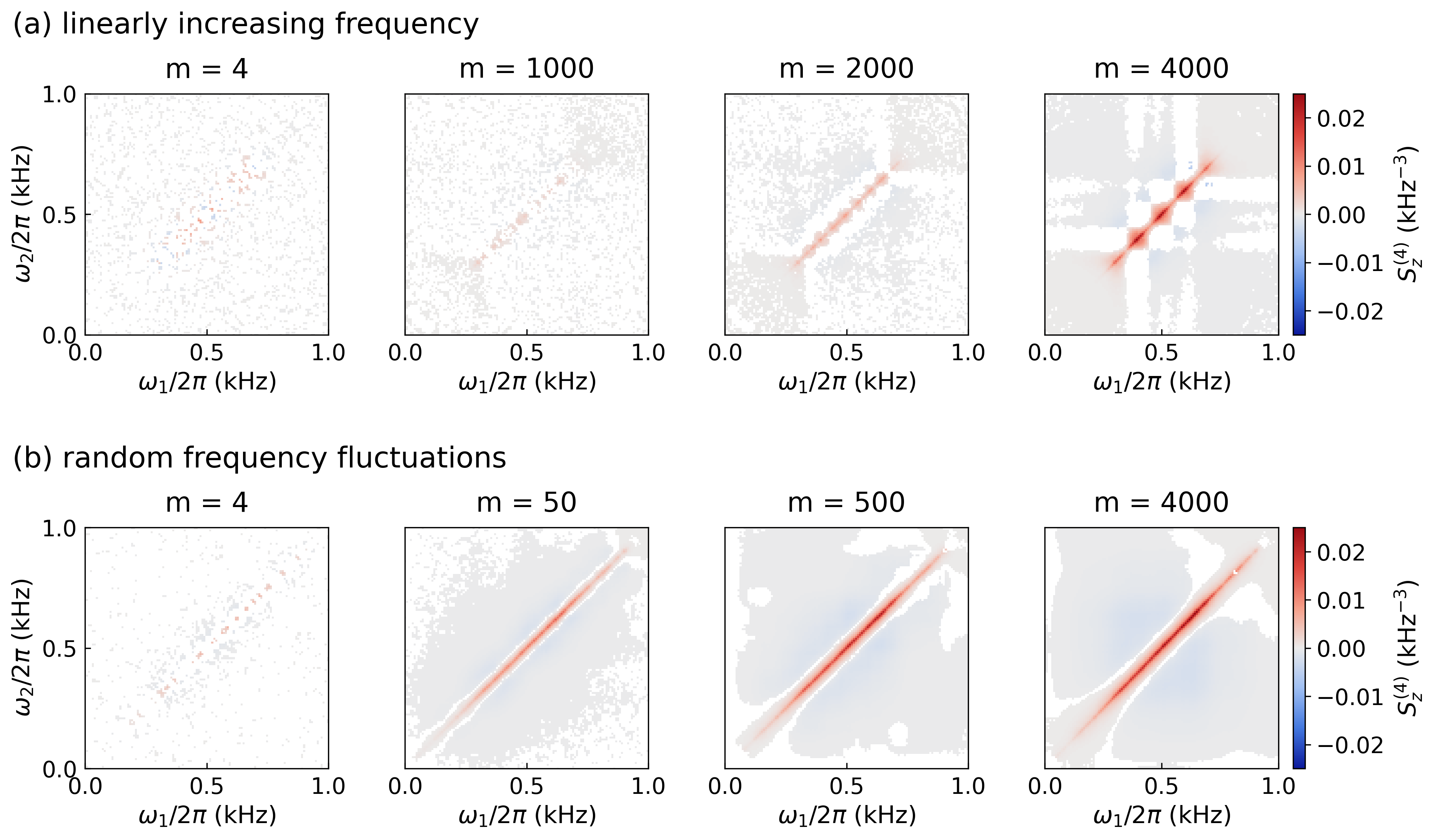}
			\caption{(a) Quasi-polyspectra $S^{(4)}$ of a non-stationary signal $x(t)$ that corresponds to a stochastically driven harmonic oscillator with a frequency that slowly drifts from $0.3$~kHz to $0.7$~kHz (see text). The spectra are an average of short-time estimates that are calculated each from $m$ sample-windows of the data. In contrast to spectra from a stationary signal, the quasi-spectra depend on $m$ and exhibit quasi-correlations that disappear only for small $m$. (b) Quasi-polyspectra $S^{(4)}$ of a stochastically driven harmonic oscillator with a slowly but randomly changing frequency in a frequency interval around $0.5$~kHz. The structures are similar to (a) and again exhibit quasi-correlations for increasing $m$. Quasi-polyspectra may serve as a tool for detecting non-stationary behavior in a signal.}     
			\label{fig:fake_correlations}   
		\end{figure*}

		Brillinger’s definition of polyspectra requires stationarity. In this section, we introduce the term “quasi-polyspectra” for spectra that were calculated from non-stationary signals with the methods presented above.  
		The stationary assumption about a signal implies that all statistical properties of the signal do not depend on time. For example, the signal $z_1(t) = A \sin(\omega t)$ is not stationary as $\langle z_1(t) \rangle = A \sin(\omega t)$ depends on time $t$. The signal $z_2(t) = \Gamma(t)$ with $\Gamma$ being white noise is stationary and fulfills e.g. $\langle z_2(t) \rangle = 0$ and 
		$\langle z_2(t) z_2(t + \tau)\rangle = \delta(\tau)$ which are independent of $t$.
		The signal $z_3(t) = \Gamma(t) + t$ is not stationary as $\langle z_3(t) \rangle = t$.
		The estimates of polyspectra along the equations presented in Section \ref{sec:HigherOrderSpectra} require in addition that the 
		Fourier coefficients $a_k$ which were calculated from subsequent windows are i.i.d. The k-statistics for estimating cumulants rests on the i.i.d. assumption (see Fisher in \cite{FisherPLMS1928}). This requirement is met approximately by processes that lose their memory of the past with a time constant of a fraction of the window length $T$. All processes presented in Section \ref{sec:HigherOrderSpectra} and \ref{sec:MultChann} fulfilled stationarity and had a short memory of the past (i.e., they exhibit a short “correlation time” – a term often used in physics). 
		
		Next, we illustrate some features of quasi-polyspectra of a non-stationary process. The signal $x(t)$ under consideration is the time-dependent position of a stochastically driven oscillator as given in Eq. \eqref{eq:linear_filter}  with a damping coefficient $\gamma = 0.8\pi$~kHz. Non-stationary behavior is introduced by linearly increasing the frequency of the oscillator from \SI{0.3}{kHz} to \SI{0.7}{kHz} over the full simulation time $T_{\rm full} = 2\times10^{3}$~s. 
		
		The quasi-power spectrum exhibits spectral weight in the frequency range from \SI{0.3}{kHz} to \SI{0.7}{kHz} in accordance with the shifting frequency (not shown). The quasi-spectrum is the sum of estimates during the time $T_{\rm full}$. Each estimate is calculated from $m$ Fourier-coefficients $a_k$ which represent a time span $m T$, where $T$ is the window length and $m$ is the number of samples used in the k-statistics for estimating cumulants.  These “short time estimates” exhibit peaks at the momentary frequency of the oscillator and are then average for yielding the quasi-spectrum with the broadened spectrum. 
		
		The process $x(t)$ is approximately Gaussian for short times, i.e., the estimates of higher-order spectra should be zero. Figure~\ref{fig:fake_correlations}(a) (left panel) shows indeed a zero quasi-trispectrum that was obtained for $m = 4$ and $T = \SI{125}{ms}$. 
		
		The features in the quasi-polyspectra change dramatically when $m$ is increased to $4000$. 
		The full spectrum is calculated from only five short-time estimates. 
		Each short-time estimate now spans a time where the frequency drift is significant. The resulting quasi-trispectra exhibit positive quasi-correlations among neighboring frequencies. Five square-like structures on the diagonal of the quasi-spectrum that resulted from the five short-time estimates are clearly visible.
		A very similar structure occurs in the quasi-polysectra when the oscillator frequency is shifted randomly 
		[see Figure~\ref{fig:fake_correlations}(b)]. For this, the frequency $f(t)$ follows the stochastic equation for an overdamped particle in a harmonic potential
		\begin{equation}
			df = - \gamma (f - f_0) \, dt + \sigma \,dW_f,
		\end{equation}
		where $\gamma = 3 \times 10^{-2}$~s$^{-1}$, $f_0 = 0.5$~kHz, and $\sigma^2 = 1.25\times10^3$~Hz$^3$. The initial condition was $f(0) = 0.5$~kHz and the simulation covered $T = 2\times10^3$~s.
		We emphasize that in the case of a stationary signal the structures of the polyspectra {\it do not depend} on the number of samples $m$.
		The changing structure of the quasi-polyspectra with $m$ is therefore a clear indication of the non-stationarity of the signal. We envision the calculation of quasi-polyspectra in dependence of $m$ as a very useful tool for detecting non-stationary behavior.
		The \textit{SignalSnap} library does in addition support storing the results of short-time estimates as sequential spectra. Those can be used to characterize non-stationary behavior in more detail.

		\section{Symmetries and equivalences of Polyspectra}
		\label{sec:symm}
		Polyspectra exhibit symmetries in a way that certain sectors in a spectrum can be mapped onto each other. Symmetries can therefore be exploited to reduce the computational cost of calculating spectra and to avoid redundancies in the display of spectra. 
		Symmetries follow from the definition of polyspectra [Eq.~\eqref{eq:defPolyspectraMulti}], where the cumulant allows for a permutation of its arguments without changing the result. In the case of real-valued channels $z_j(t)$, the relation $z_j(-\omega) = z^*_j(\omega)$ further increases the symmetry in a spectrum.
		
		The {\it second-order spectra} fulfill
		\begin{equation}
			S^{(2)}_{xy} (\omega) =  [S^{(2)}_{xy} (-\omega)]^*,
		\end{equation}
		which follows from 
		\begin{align}
			C_2(x(\omega_1),y(\omega_2)) & =  C_2(x^*(\omega_1),y^*(\omega_2))^* \nonumber \\
			& =  C_2(x(-\omega_1),y(-\omega_2))^*.
		\end{align} 
		Moreover,
		\begin{equation}
			S^{(2)}_{xy} (\omega) =  S^{(2)}_{yx} (-\omega) = [S^{(2)}_{yx} (\omega)]^*, \label{S2symm}
		\end{equation}
		holds, because of $\omega_1 + \omega_2 = 0$ and
		\begin{align}
			C_2(x(\omega_1),y(\omega_2)) & =  C_2(y(\omega_2),x(\omega_1)) \nonumber \\
			& =  C_2(y(-\omega_1),x(-\omega_2)).
		\end{align} 
		Equation (\ref{S2symm}) shows that the spectra $S^{(2)}_{xy}$ and $S^{(2)}_{yx}$ are equivalent. The well known symmetry for one channel, $S^{(2)}_{zz}(\omega) = S^{(2)}_{zz}(-\omega)$, is recovered from Eq. (\ref{S2symm}). 
		The higher-order spectra $S^{(3)}$ and $S^{(4)}$
		exhibit many more equivalencies between spectra that we discuss in the following.
		
\begin{figure*}[t]
			\centering
			\includegraphics[width=\textwidth]{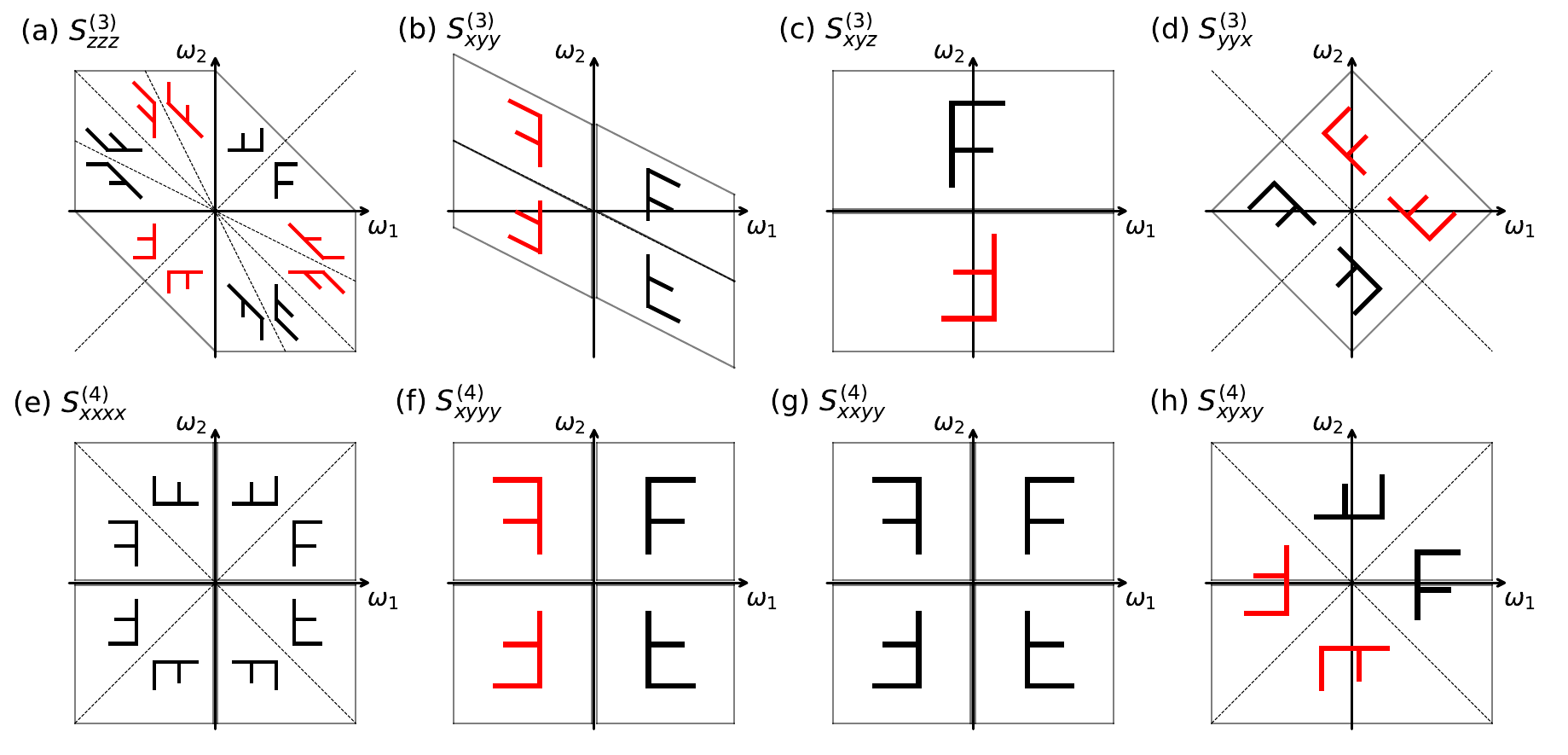}   
			\caption{Symmetries of polyspectra $S^{(3)}$ and $S^{(4)}$ regarding their dependency on $\omega_1$ and $\omega_2$. The red and black "F" labels sectors where the spectral values are the complex conjugate of each other. Spectra (b) and (d) are equivalent and can be mapped onto each other.}
			\label{fig:poly_sym}
		\end{figure*}

		We begin by investigating the symmetries of the {\it third-order spectra}.  In case of one channel $z(t)$, the spectrum $S^{(3)}_{zzz}(\omega_1,\omega_2)$
		fulfills the relation 
		\begin{equation}
			S^{(3)}_{zzz}(\vec{\omega}’) = S^{(3)}_{zzz}(\vec{\omega}),\,\,\,\vec{\omega}’ = T_j \vec{\omega}
		\end{equation}
		trivially for $T_1 = \left(\begin{array}{cc} 1 & 0 \\ 0 & 1 \end{array} \right)$.
		It also holds for $T_2 =  \left( \begin{array}{cc} 0 & 1 \\ 1 & 0 \end{array} \right)$ because
		\begin{equation}
			C_3(z(\omega_1),z(\omega_2), z(\omega_3))  = C_3(z(\omega_2),z(\omega_1), z(\omega_3)).  
		\end{equation}
		Another transformation $T_3 =  \left( \begin{array}{cc} 1 & 0 \\ -1 & -1 \end{array} \right)$
		holds, because
		\begin{equation}
			C_3(z(\omega_1),z(\omega_2), z(\omega_3))  = C_3(z(\omega_1),z(-\omega_1-\omega_2), z(\omega_2)),  
		\end{equation}
		where we used $\omega_1 + \omega_2 + \omega_3 = 0$.
		Sequential applications of $T_1$, $T_2$, and $T_3$ leads to exactly three new transformations $T_4 = T_2 T_3$, $T_5 = T_3 T_2$, and $T_6 = T_3 T_2 T_3$. 
		The spectrum therefore can be divided into six equivalent sectors. Regarding, that for real-valued $z(t)$ we have
		$S^{(3)}_{zzz}(\vec{\omega}) = [S^{(3)}_{zzz}(-\vec{\omega})]^*$, another transformation
		$T_7 =  \left( \begin{array}{cc} -1 & 0 \\  0 & -1 \end{array} \right)$ is possible that maps the complex conjugate values of one part of the spectrum 
		to another sector.
		We therefore end up with 12 transformations that divide the spectrum into 12 equivalent sectors (also noted in \cite{rosenblatt1965estimation}).
		Figure \ref{fig:poly_sym}(a) shows all twelve sectors and their symmetry relations. Regions containing the symbol "F" in different colors contain the 
		mutually complex conjugate values of the spectrum. 
		
		The spectrum for two channels $S^{(3)}_{xyy}(\vec{\omega})$ shows a reduced symmetry since only $T_1$, $T_3$ and $T_7$ 
		as well as  $T' = T_3 T_7$ can be applied. Consequently, $S^{(3)}_{xyy}(\vec{\omega})$ can be divided into four equivalent sectors [Figure \ref{fig:poly_sym}(b)].  The special shape of the sectors is consistent with points on a line $(t, -t/2)$ [for $t$ real] that keep their position under transformation with $T_3$. The special shape also explains the structure of the spectrum $S^{(3)}_{uxx}$ displayed in Fig. \ref{fig:multi_det_S3_S4}. 
		The spectrum $S^{(3)}_{yyx}(\vec{\omega})$ exhibits a symmetry with regard to
		$T_1$, $T_2$, and $T_7$ if $x$ and $y$ are real. The spectrum can then be divided into 4 equivalent sectors [Figure \ref{fig:poly_sym}(d)]. 
		It is important to note that only one of the spectra  $S^{(3)}_{xyy}(\vec{\omega})$  and  $S^{(3)}_{yyx}(\vec{\omega})$ has to be calculated as they contain the same information.
		Their relation of equivalence
		\begin{equation}
			S^{(3)}_{yyx}(\vec{\omega}) = S^{(3)}_{xyy}(\vec{\omega}')\,\,\,\mbox{with}\,\,\vec{\omega}' = \left( \begin{array}{cc} -1 & -1 \\ 1 & 0 \end{array} \right)\vec{\omega} 
		\end{equation}
		is easily shown from the cumulant definition of polyspectra.
		
		The spectrum for three channels  $S^{(3)}_{xyz}(\vec{\omega})$ is the complex conjugate of itself under transformation with $T_7$ and exhibits two sectors [Figure \ref{fig:poly_sym}(c)].
		
		Next, we discuss the symmetry properties of two-dimensional cuts of the {\it fourth-order spectra}. The spectrum $S^{(4)}_{zzzz}(\omega_1,\omega_2)$ follows from the cumulant
		$C_4(z(\omega_1),z(-\omega_1),z(\omega_2),z(-\omega_2)$. 
		The spectrum is always real for real-valued $z(t)$ since
		\begin{align}
			C_4(z(\omega_1),z(-\omega_1),z(\omega_2),z(-\omega_2) \hspace{-3cm} &   \nonumber \\ 
			& =             (C_4(z(\omega_1)^*,z(-\omega_1)^*,z(\omega_2)^*,z(-\omega_2)^*)^* \nonumber \\
			& =  C_4(z(-\omega_1),z(\omega_1),z(-\omega_2),z(\omega_2))^* \nonumber \\
			& =  C_4(z(\omega_1),z(-\omega_1),z(\omega_2),z(-\omega_2))^*,
		\end{align}
		where we used $z(\omega)^* = z(-\omega)$ in the second line, and equivalence of cumulants under permutation of arguments in the last line. 
		
		Similarly to the procedure in the third-order case, we find identical spectra $S^{(4)}_{zzzz}(\vec{\omega}’) =
		S^{(4)}_{zzzz}(\vec{\omega})$ for $T_1, T_2, T_7$, and $T_{13} =  \left( \begin{array}{cc} 1 & 0 \\ 0 & -1 \end{array} \right)$. Four other transformations follow from sequentially applying $T_1, T_2, T_7$, and $T_{13}$. The spectrum $S^{(4)}_{zzzz}(\vec{\omega})$ can therefore be divided into 8 equivalent sectors [see Figure \ref{fig:poly_sym}(e)]. 
		The spectrum $S^{(4)}_{xyyy}(\vec{\omega})$ is identical under transformation with $T_1$, and $T_{13}$.
		Transformation with $T_7$ yields the complex conjugate of the spectrum. A combination of all transformations leads to 4 equivalent sectors
		[see Figure \ref{fig:poly_sym}(f)]. 
		Similarly, we find $S^{(4)}_{xxyy}(\vec{\omega})$ to be real and equivalent in 4 sectors [see Figure \ref{fig:poly_sym}(g)].
		The spectrum   $S^{(4)}_{xyxy}(\vec{\omega})$ is in general complex-valued and is equivalent in four sectors [see Figure \ref{fig:poly_sym}(h)].
		
		There are a number of other fourth-order spectra that are equivalent to $S^{(4)}_{xyyy}(\vec{\omega})$ and $S^{(4)}_{xyxy}(\vec{\omega})$.
		The two-channel spectra $S^{(4)}_{xyyy}(\vec{\omega})$, $S^{(4)}_{yxyy}(\vec{\omega})$,  $S^{(4)}_{yyxy}(\vec{\omega})$, $S^{(4)}_{yyyx}$ are equivalent under appropriate transformation of $\vec{\omega}$ (not shown). The spectrum   $S^{(4)}_{xyxy}(\vec{\omega})$ is equivalent to
		$S^{(4)}_{xyyx}(\vec{\omega})$. 
		
		We briefly mention that in the three-channel case only the spectra $S^{(4)}_{xyzz}(\omega_1,\omega_2)$ and   $S^{(4)}_{xzyz}(\omega_1,\omega_2)$  
		need to be calculated separately. The four-channel case requires the calculation of  $S^{(4)}_{xyzw}(\omega_1,\omega_2)$,
		$S^{(4)}_{xzyw}(\omega_1,\omega_2)$  and  $S^{(4)}_{xwyz}(\omega_1,\omega_2)$. All other permutations of channels yield equivalent spectra. 
		Symmetries of three-dimensional single-channel spectra $S^{(4)}_{z}(\omega_1,\omega_2,\omega_3)$ have been discussed in \cite{pflugJASA1992}.

		\section{Advancing the SignalSnap library}
		The task of calculating polyspectra following the equations of Sections II, III, and V allows for parallelization. SignalSnap exploits this
		fact by computing the values of a spectrum at different frequency-pairs $(\omega_1, \omega_2)$ in parallel. 
		At present SignalSnap makes heavy use of numerics with three-dimensional arrays. 
		Two dimensions are used for the frequency dependent products of Fourier coefficients and another dimension is used for the coefficients from $m$ subsequent temporal windows. In that way a single short-time estimate for the spectra $S^{(3)}$ and  $S^{(4)}$ can be calculated efficiently. SignalSnap is built on the ArrayFire library which provides powerful operations for manipulating arrays on either CPUs or GPUs \cite{Yalamanchili2015}. 
		While the use of GPUs in SignalSnap provides an extreme speedup in comparison 
		with calculations on a CPU, future implementations may become even faster by addressing the following points.
		(i) At present SignalSnap evaluates only a 
		single short-time estimate at a time, before more data is streamed on the GPU for processing. In the case of less demanding spectra with a moderate number of points and a small number $m$ of windows, the computation may require
		only a fraction of the computational capacity of the GPU. A future version of SignalSnap could instead compute several short-time estimates of the spectra in parallel making e.g. use of a four-dimensional array with a fourth dimension 
		for subsequent estimates. In such a way, the overall rate of data and therefore the spectral bandwidth for, e.g., real-time evaluation of spectra could be increased to the ultimate limit set by the GPU hardware. Currently, increasing $m$ or the window length $T$ along with $N$ is a way to increase the GPU load and process more data at once.
		(ii) The parallel computation of $S^{(2)}$, $S^{(3)}$, and  $S^{(4)}$ 
		currently leads to many identical intermediate results, thus the same computation is done more than once. A speedup may be obtained by a cleverer way of using intermediate results for calculating all spectra. (iii) We have begun transitioning from the ArrayFire library to PyTorch, with initial results already demonstrating significant speed improvements in calculations. Libraries such as PyTorch and JAX are widely adopted within the machine learning community and offer straightforward installation processes. We also expect them to be constantly updated for new hardware. We also have not experimented with the optimal use of software libraries. Operations on multidimensional arrays may be fast for one dimension but slower for another. Taking this into account in an implementation could yield further speedup. Future implementations of SignalSnap may even test the available hardware themselves to decide on the best algorithm which is a technique used by the celebrated FFTW implementation of the fast Fourier transformation \cite{frigoIEEE2005}.     
		
		Apart from improving numerics, the identification and implementation of estimators with improved statistics may further improve the appeal of SignalSnap (see Paragraph \ref{para:unbiasedestimators}),
		
		\section{Conclusion}
		In conclusion, we have introduced unbiased and consistent estimators for estimating single- and multi-channel polyspectra from real-world data. We established a clear relationship between the definition of ideal spectra and their estimates, which is essential for comparing theoretical spectra with their measured counterparts (see, e.g., \cite{sifftPRR2021}). We also presented example polyspectra calculated using our GPU-based SignalSnap library, demonstrating that the analysis of large datasets is feasible. Our work significantly lowers the barriers to adopting polyspectral analysis, enabling broader application among experimentalists and engineers.
		Moreover, this foundation may inspire statisticians to develop optimal k-statistic-based estimators for spectral analysis, ultimately enhancing the accuracy and applicability of polyspectral methods.

		\section*{Acknowledgment}
		We acknowledge financial support by the Deutsche Forschungsgemeinschaft (DFG) under Project No. 510607185 as well as by the Mercator Research Center Ruhr (MERCUR) under Project No. Ko-2022-0013.

		\appendix
		\section{Conventions for Fourier Transformations and Convolution Integrals}
		\label{app:Fourier}
		The function \( f(t) \) and its Fourier transform \( f(\omega) \) are only distinguished by their arguments. They are related via
		\begin{align}
			f(\omega) &= \int_{-\infty}^{\infty} e^{j \omega t} f(t) \,dt,  \\
			f(t) &= \frac{1}{2\pi} \int_{-\infty}^{\infty} e^{-j \omega t} f(\omega) \,d\omega. 
		\end{align}
		The convolution in time is defined as
		\begin{equation}
			f(t) \ast g(t) = \int_{-\infty}^\infty f(t - \tau) g(\tau) \,d\tau
		\end{equation}
		and the convolution in frequency as
		\begin{equation}
			\label{eq:convolution}
			f(\omega) \ast g(\omega) = \frac{1}{2\pi} \int_{-\infty}^\infty f(\omega - \nu) g(\nu) \, d\nu, 
		\end{equation}
		with the additional prefactor \((2\pi)^{-1}\). This results in the following relations for the Fourier transforms of convolutions and products of functions:
		\begin{align}
			h_1(t) &= f(t) \ast g(t), \\
			h_1(\omega) &= f(\omega) g(\omega), 
		\end{align}
		and
		\begin{align}
			h_2(\omega) &= f(\omega) \ast g(\omega),  \\
			h_2(t) &= f(t) g(t). 
		\end{align}
		\section{Approximate Confined Gaussian Window}
		\label{app:cgw}
		The discrete approximate confined Gaussian window is defined as \cite{starosielecSP2014}
		\begin{align}
			g_k^{(\mathrm{acG})} \propto G(k)-G(-1 / 2) \frac{G(k+N)+G(k-N)}{G(-1 / 2+N)+G(-1 / 2-N)} \nonumber \\
			\label{eq:confined_gaussian}
		\end{align}
		for $k=0,1,...,N-1$ 
		with the Gaussian function $G(x)=\exp \left[-(x-(N- 1) / 2)^2 / (4 N^2 \sigma_t^2) \right]$ centred at $(N-1)/2$.
		The widths of the Gaussian is set in SignalSnap to a default value $\sigma_t = 0.14$. 
		\section{Spectra of processes with time inversion symmetry}
		\label{app:timeinversion}
		Consider the process $z(t)$ and its time inversion $y(t) =  z(-t)$. In general their spectra
		have the relation $S^{(n)}_z = (S^{(n)}_y)^*$ which follows from
		$y(\omega) = \int y(t) e^{j \omega t} \,dt
		\allowbreak = \int z(-t) e^{j \omega t} \,dt
		\allowbreak = \int z(t') e^{-j \omega t'} \,dt'
		\allowbreak = z(-\omega)$ and
		\begin{align}
			C_n(y(\omega_1), y(\omega_2), …) & =  C_n(z(-\omega_1), z(-\omega_2), …) \nonumber \\
			& = C_n(z^*(\omega_1), z^*(\omega_2), …) \nonumber \\
			& =  \left[ C_n(z(\omega_1), z(\omega_2), …)\right]^*. \nonumber 
		\end{align}
		In case of time-inversion symmetry, where $z(t)$ and $z(-t)$ have same statistics, $S^{(n)}_z = (S^{(n)}_z)^*$ follows \cite{efimovQJRMS2001}. Consequently, any 
		signal with time inversion symmetry gives rise to real-valued spectra. Any spectrum with a non-vanishing imaginary part proves a violation of time-inversion symmetry, but not the other way around. The above proof holds analogously for spectra of multi-channel processes.
		
		\section{Polyspectra of complex-valued signals}
		\label{app:polyspeccomplex}
		The definition of polyspectra, Eq. (\ref{eq:defPolyspectraMulti}), includes spectra of complex-valued signals, while SignalSnap is presently restricted to the case of real-valued signals. This limitation can be circumvented by exploiting the multilinearity of polyspectra. We demonstrate this for the case of a cross-correlation spectrum of two complex-valued signals $z_1(t) = x_1(t) + j y_1(t)$ and $z_1(t) = x_1(t) + j y_1(t)$, where $x_i(t)$ and $y_i(t)$ are real. 
		Using multilinearity, we find
		\begin{align}
			S^{(2)}_{z_1 z_2} = S^{(2)}_{x_1 x_2} + j S^{(2)}_{x_1 y_2} + j S^{(2)}_{y_1 x_2} - S^{(2)}_{y_1 y_2},
		\end{align} 
		where the four spectra on the right side are from real-valued signals which can be estimated by SignalSnap. Similarly, third and fourth-order spectra of complex-valued signals can be calculated which result in a sum of 8 or 16 real-valued spectra, respectively.

		\section{The full three-dimensional fourth-order spectrum}
		\label{app:S4threeD}
		SignalSnap implements presently a two-dimensional cut 
		\begin{equation}
			S^{(4)}(\omega_1,\omega_2) = S^{(4)}(\omega_1,-\omega_1,\omega_2)
		\end{equation}
		through the full three-dimensional fourth-order spectrum. 
		Noting that  $S_{z_1 z_2 z_3 z_4}^{(4)}(\omega_1,\omega_2)$ is related to 
		$C_4(z_1(\omega_1),\allowbreak z_2(-\omega_1),\allowbreak z_3(\omega_2),\allowbreak z_4(-\omega_2))$, we find that a relation to a general
		$C_4(z_1(\omega_1 + \Delta \omega),\allowbreak z_2(-\omega_1+ \Delta \omega),\allowbreak z_3(\omega_2 – \Delta \omega),\allowbreak z_4(-\omega_2 -  \Delta \omega))$ can be established via
		\begin{align}
			x(t) & =  z_1(t) e^{j \Delta \omega t} \nonumber \\
			y(t) & =  z_2(t) e^{-j \Delta \omega t} \nonumber \\
			z(t) & =  z_3(t) e^{-j \Delta \omega t} \nonumber \\
			w(t) & =  z_4(t) e^{j \Delta \omega t} \nonumber 
		\end{align}
		which yields
		\begin{equation}
			S^{(4)}_{z_1 z_2 z_3 z_4}(\omega_1 + \Delta \omega, -\omega_1 + \Delta \omega, \omega_2 - \Delta \omega) =   S^{(4)}_{xyzw}(\omega_1,\omega_2).
		\end{equation}
		Consequently, all planes parallel to the one provided by  
		\begin{equation}
			S^{(4)}_{z_1 z_2 z_3 z_4}(\omega_1,\omega_2)
		\end{equation} 
		can be obtained by introducing frequency shifted version of $z_j$ via the above definitions for $x(t)$, $y(t)$, $z(t)$, and $w(t)$. The newly defined signals are in general complex, i.e. the spectrum
		on the right-hand side must be evaluated with the method of \ref{app:polyspeccomplex}.

		\section{Estimators of the third- and fourth-order spectrum}
		\label{C3S3}
		
		Starting from Brillinger's definition,
		\begin{equation}
			C_3\Bigl(z(\omega), z(\omega'), z^*(\omega'')\Bigr) 
			= 2\pi\, \delta(\omega + \omega' - \omega'')\, S^{(3)}_z(\omega,\omega'),
		\end{equation}
		we find for the third-order cumulant of Fourier coefficients
		\begin{align}
			C_3(a_k,a_l,a^*_{k+l}) &\approx C_3(a'_k,a'_l,a'^*_{k+l}) \nonumber\\[1mm]
			& \hspace{-0.5cm} = \frac{1}{(2\pi)^3} \iiint 
			C_3\Bigl(z(\omega), z(\omega'), z^*(\omega'')\Bigr)\nonumber\\
			& \hspace{-0.5cm}\quad \times g(\omega_k-\omega) \, g(\omega_k-\omega') \nonumber\\[1mm]
			& \hspace{-0.5cm}\quad \times g^*(\omega_{k+l}-\omega'') \, d\omega\, d\omega'\, d\omega'' \nonumber\\[1mm]
			& \hspace{-0.5cm}= \frac{1}{(2\pi)^2} \iint 
			S^{(3)}_z(\omega,\omega')\, g(\omega_k-\omega)\, g(\omega_l-\omega') \nonumber\\[1mm]
			&\hspace{-0.5cm} \quad \times g^*\Bigl(\omega_{k+l}-(\omega+\omega')\Bigr)
			\, d\omega\, d\omega'.
		\end{align}
		
		The ideal spectrum $S^{(3)}_z(\omega,\omega')$ is convoluted with the 
		two-dimensional function
		\begin{equation}
			\tilde{g}(\omega,\omega') = g(\omega)\, g(\omega')\, g^*(\omega+\omega'),
		\end{equation}
		whose integral is
		\begin{align}
			A^{(3)} &= \iint g(\omega)\, g(\omega')\, g^*(\omega+\omega')\, d\omega\, d\omega' \nonumber\\[1mm]
			&= \int\!\cdots\!\int g(t)\, g(t')\, g^*(t'')\,
			e^{\,j\omega t + j\omega' t' - j(\omega+\omega')t''} \nonumber\\[1mm]
			&\quad \times dt\, dt'\, dt''\, d\omega\, d\omega' \nonumber\\[1mm]
			&= (2\pi)^2 \int g(t)\, g(t)\, g^*(t)\, dt \nonumber\\[1mm]
			&\approx (2\pi)^2 \frac{T}{N} \sum_{i=0}^{N-1} g_i^2\, g^*_i.
		\end{align}
		
		An analog calculation can be performed for the fourth-order spectrum, leading to 
		\begin{align}
			C_4(a_k,a_l,a_p,a^*_{k+l+p}) &\approx   \nonumber \\
			& \hspace{-2.5cm}= \frac{1}{(2 \pi)^3}  \iiint_{-\infty}^{\infty}  S^{(4)}_z(\omega,\omega', \omega'') \nonumber\\
			& \hspace{-2.5cm}\times g(\omega_k - \omega)  g(\omega_l - \omega') g(\omega_p - \omega'') \nonumber \\
			& \hspace{-2.5cm}\times g^*(\omega_{k+l+p} - (\omega+\omega'+\omega'')) \,d\omega  \,d \omega' \,d \omega''.
		\end{align}
		The ideal spectrum $S^{(4)}_z(\omega,\omega', \omega'')$ is convoluted with the three-dimensional function 
		\begin{equation}
			\tilde{\tilde{g}}(\omega, \omega', \omega'') =  g( \omega)  g( \omega') g( \omega'') g^*(\omega+\omega'+\omega''),
		\end{equation}
		whose integral is
		\begin{align}
			A^{(4)} & \approx  (2 \pi)^3 \frac{T}{N} \sum_{j=0}^{N-1} g_j g_j g_j g^*_j.
		\end{align}

		\section{Estimator of the first-order spectrum}
		\label{C1S1}
		We find for the first-order cumulant:
		\begin{align}
			C_1(a_k) & \approx  C_1(a'_k) \nonumber \\
			& =  \frac{1}{2 \pi} \int  C_1(z(\omega)) g(\omega_k - \omega) \,d\omega\,
			\nonumber \\
			& =   S^{(1)}_z g(\omega_k) \nonumber \\
			& =  S^{(1)}_z \int g(t) e^{j \omega_k t}\, dt \nonumber \\
			& \approx  S^{(1)}_z  \frac{T}{N} \sum_{i=0}^{N-1} g_i,
		\end{align}
		where we assumed $k = 0$ in the last line.
		We obtain
		\begin{equation}
			S_z^{(1)}  \approx \frac{N C_1(a_0)}{T \sum_{i = 0}^{N-1} g_i}. 
		\end{equation}
		
		\section{Fourth-order cumulant estimator}
		\label{app:factorizedC4}
		The fourth-order cumulant estimator introduced in the main text (Eq.~\ref{eq:c4_estimator}) can be factorized as
		\begin{align}
			c_4^{(\mathrm{a})}(x, y, z, w) \hspace{-1cm} &  \nonumber \\
			& =  \frac{m^2}{(m-1)(m-2)(m-3)} \nonumber\\
			&  \times[(m+1)(\overline{x-\bar{x})(y-\bar{y})(z-\bar{z})(w-\bar{w})}\nonumber \\
			&  -(m-1)(\overline{(x-\bar{x})(y-\bar{y})} \times \overline{(z-\bar{z})(w-\bar{w})}   \nonumber \\
			&  + \overline{(x-\bar{x})(z-\bar{z})} \times \overline{(y-\bar{y})(w-\bar{w})}   \nonumber \\
			&  + \overline{(x-\bar{x})(w-\bar{w})} \times \overline{(y-\bar{y})(z-\bar{z})}   )]. 	\label{eq:factorizedC4}			
		\end{align}
		where $\overline{\cdots}$ denotes the average over $m$ samples, $\bar{x}$ represents the mean of the variable $x$. The factorization reduces the number of necessary multiplications and speeds up the calculation of the cumulant. In our SignalSnap library the factorized form is implemented.
		
		\section{Realization of Lorentzian Bandpass Filtered White Noise}
		\label{app:bandpass}
		We generated Lorentzian bandpass-filtered white noise by integrating the differential equation
		\begin{equation}
			\frac{dy_c(t)}{dt} = \Bigl(j\,\omega_1 - \gamma_1\Bigr)y_c(t) + \gamma_1\,\Gamma(t)
		\end{equation}
		with the center angular frequency $\omega_1$ and the parameter $\gamma_1$ that sets the bandwidth of the filter. Here, \(\Gamma(t)\) represents white noise process with $\langle \Gamma(t)^*\Gamma(t')\rangle = \delta(t-t')$.  The integration results in a complex-valued output signal $y_c(t)$.
		
		The transfer function associated with this filter exhibits a Lorentzian magnitude response. The response is given by
		\[
		|H(\omega)|^2 = \frac{1}{1+\Bigl(\frac{\omega-\omega_1}{\gamma_1}\Bigr)^2}. \]
		The examples presented in the main text utilize only the real part of the complex signal $y(t) = \mathfrak{Re}\left(y_c(t)\right)$ as the physical signal.

		\bibliographystyle{elsarticle-num} 
		\bibliography{spinnoiseV2}
		
		
		
		
		
		
\end{document}